# Inhomogeneous Reionization Models in Cosmological Hydrodynamical Simulations

Jose Oñorbe,[1,2][★] F. B. Davies,[3,2] Z. Lukić,[4] J. F. Hennawi[3,2] and D. Sorini[2,5]
[1]*Institute for Astronomy, University of Edinburgh, Blackford Hill, EH9 3HJ, Edinburgh, United Kingdom*
[2]*Max-Planck-Institut für Astronomie, Königstuhl 17, 69117 Heidelberg, Germany*
[3]*Department of Physics, University of California, Santa Barbara, CA 93106-9530, USA*
[4]*Lawrence Berkeley National Laboratory, CA 94720-8139, USA*
[5]*Fellow of the International Max Planck Research School for Astronomy and Cosmic Physics at the University of Heidelberg (IMPRS-HD)*

**ABSTRACT**
In this work we present a new hybrid method to simulate the thermal effects of reionization in cosmological hydrodynamical simulations. The method improves upon the standard approach used in simulations of the intergalactic medium (IGM) and galaxy formation without a significant increase of the computational cost, thereby allowing for efficient exploration of the parameter space. The method uses a small set of phenomenological input parameters, and combines a seminumerical reionization model to solve for the topology of reionization with an approximate model of how reionization heats the IGM using the massively parallel Nyx hydrodynamics code which is specifically designed to solve for the structure of diffuse IGM gas. We have produced several medium-scale high-resolution simulations ($2048^3$, $L_{\rm box}$ = 40 Mpc h$^{-1}$) with different instantaneous and inhomogeneous H I reionization models that use this new methodology. We study the IGM thermal properties of these models and find that large scale temperature fluctuations extend well beyond the end of reionization. By analyzing the 1D flux power spectrum of these models, we find up to $\sim$ 50% differences in the large-scale properties (low modes, $k \lesssim 0.01$ s km$^{-1}$) of the post-reionization power spectrum as a result of the thermal fluctuations. We show that these differences could allow one to distinguish between different reionization scenarios with existing Ly$\alpha$ forest measurements. Finally, we explore the differences in the small-scale cut-off of the power spectrum finding that, for the same heat input, models show very good agreement provided that the reionization redshift of the instantaneous reionization model occurs at the midpoint of the inhomogeneous model.

## 1 INTRODUCTION

The reionization of neutral hydrogen (H I) in the Universe is believed to be driven by the ultraviolet radiation emitted by the first galaxies and quasars. Our current knowledge about H I reionization comes mainly from two observational probes: an integral constraint from the electron scattering optical depth of the cosmic microwave background (CMB) and the measured transmission of the Lyman-$\alpha$ (Ly$\alpha$) forest (McQuinn 2016; Dayal & Ferrara 2018).

The most recent Planck satellite measurements of the CMB Thomson scattering optical depth, $\tau_e$, imply that hydrogen was reionized by $z \sim 7.3 - 10.6$ ($\sim 2\sigma$ limits from Planck Collaboration et al. 2018). However, several observations seem to indicate that the end of reionization occurs at $z \sim 6$: the complete Gunn-Peterson absorption (Gunn & Peterson 1965) measured in high-$z$ quasar spectra and the the rapid increase of the Ly$\alpha$ optical depth and its scatter (Fan et al. 2006; Becker et al. 2015; Bosman et al. 2018; Eilers et al. 2018). At higher redshifts $z \gtrsim 6$, the Ly$\alpha$ opacity can only set lower limits on the redshift of reionization, but measurements of the damping wing of H I in the intergalactic medium (IGM) in the spectra of $z > 7$ quasars have recently started to provide interesting constraints on the hydrogen neutral fraction at these redshifts consistent with CMB observations (Mortlock et al. 2011; Greig et al. 2017; Davies et al. 2018b; Bañados et al. 2018). Therefore the exact details of hydrogen reionization, for example the main sources responsible for the ionizing photons, are still not known and in fact several large observational projects within the astrophysics community are underway, with the goal of finally fully understanding this fundamental cosmological process (e.g. JWST, SKA, HERA).

A key aspect of the reionization of H I is that it plays a crucial role in the thermal history of the Universe. During H I reionization, ionization fronts around the first sources propagate supersonically

[★] E-mail: onorbe@roe.ac.uk

and the IGM gas that has been cooling since the Big Bang is photoheated. Studies have suggested a post-reionization temperature of order $2 \times 10^4$ K depending on the spectral shape of ionizing sources and the speed of ionization fronts (Miralda-Escudé & Rees 1994; Abel & Haehnelt 1999; Tittley & Meiksin 2007; McQuinn 2012; Davies et al. 2016; Park et al. 2016; Finlator et al. 2018; D'Aloisio et al. 2018a). After reionization, hydrodynamical simulations show that the IGM expands and cools, but because cooling times in the low-density IGM are long, the thermal memory of this heat injection can persist for hundreds of Myr ($\Delta z = 1$) (Trac et al. 2008; Furlanetto & Oh 2009). Thus if the temperature of the IGM can be measured just after reionization at $z \sim 5 - 6$, before the thermal state has fully relaxed, the nature and timing of reionization can be constrained (Cen et al. 2009; Lidz & Malloy 2014; Nasir et al. 2016; Oñorbe et al. 2017a,b; Boera et al. 2018). The temperature of the IGM is not the entire story, however. Although baryons in the IGM trace dark matter fluctuations on Mpc scales, on smaller scales ($\sim 100$ kpc) the gas is pressure-supported against gravitational collapse by its finite temperature ($T \sim 10^4$ K, Gnedin & Hui 1998a; Kulkarni et al. 2015; Oñorbe et al. 2017a,b). Analogous to the classic Jeans argument, baryonic fluctuations are suppressed relative to the pressureless dark matter (which can collapse). The pressure smoothing scale of the IGM thus provides an integrated record of the thermal history of the IGM, and is sensitive to the timing of and heat injection by reionization events (Gnedin & Hui 1998a; Kulkarni et al. 2015; Nasir et al. 2016; Oñorbe et al. 2017a,b).

The latest measurements of the Ly$\alpha$ optical depth at high redshift have found enhanced scatter at $z > 5.5$ that cannot be explained by density fluctuations alone (Fan et al. 2006; Becker et al. 2015; Bosman et al. 2018; Eilers et al. 2018). The origin of these fluctuations has, however, been hotly debated. One possibility is that the origin lies in spatial variations in the radiation field (Davies & Furlanetto 2016; Gnedin et al. 2017; D'Aloisio et al. 2018b) or in the temperature field (D'Aloisio et al. 2015; Davies et al. 2018a), both types of which may result from a patchy, extended, and late-ending reionization process. The existence of these fluctuations could indicate that rather than the faintest galaxies (Robertson et al. 2015), rare active galactic nuclei (AGNs) play a fundamental role in reionization (Chardin et al. 2015, 2017, but see D'Aloisio et al. 2017), which is intriguing given that a large population of faint $z \sim 6$ AGNs may have just been discovered (Giallongo et al. 2015, see however Weigel et al. 2015).

Thus, previous work has demonstrated that there is not a sharp transition from the epoch of reionization to the simple uniform UV background and tight temperature-density relationship that govern the Ly$\alpha$ forest at lower redshifts $z < 5$. Instead, spatially coherent radiation and temperature fluctuations are expected to persist even after reionization is complete, leaving observable artefacts in the post-reionization IGM. As a result, the study of the statistical properties of the $z \sim 5 - 6$ IGM, where the transmitted flux is still significant, can address fundamental questions about reionization. In fact, current Ly$\alpha$ forest measurements at $z > 4$ (Viel et al. 2013a; Iršič et al. 2017; D'Aloisio et al. 2018b) provide constraints on the power at the relevant scales ($10^{-3} > k > 10^{-1}$ s km$^{-1}$, $10^{-1} > k > 10$ h/Mpc) and therefore may already tell us something about the effects of inhomogeneous reionization on the IGM at these redshifts. Moreover the past several years have seen a dramatic fivefold increase in the number of $z \gtrsim 6$ quasars from deep wide-field optical/IR surveys (Willott et al. 2010; Bañados et al. 2014; Venemans et al. 2015b,a; Wang et al. 2017). The ongoing spectroscopic follow-up of these objects (e.g. Eilers et al. 2018) ensures a significant precision boost of Ly$\alpha$ forest measurements at these redshifts in the near future.

Despite the apparent simplicity, accurate simulations of the IGM are hard because of the presence of two spatial scales. High spatial resolution ($\Delta x \lesssim 20$ kpc) is required to resolve the density structure of the IGM, while a relatively large box size ($L_{box} \gtrsim 100$ Mpc) is required both to capture the large-scale power and to obtain a fair sample of the Universe to capture the reionization fluctuation scale (see e.g. Iliev et al. 2014; Lukić et al. 2015; Oñorbe et al. 2017b; Bolton et al. 2017). Ideally it would be possible to run radiative transfer hydrodynamical simulations that are able to provide an accurate description of the sources of ionizing photons (stars, quasars, etc.). Although significant progress on this front has been made (e.g. Meiksin & Tittley 2012; Wise et al. 2014; So et al. 2014; Gnedin 2014; Pawlik et al. 2015; Norman et al. 2015; Ocvirk et al. 2016; Kaurov 2016; La Plante et al. 2017), these simulations are still too costly for exploration of the parameter space. Postprocessing radiative transfer approaches have also been used in the field (e.g. Trac & Cen 2007; Iliev et al. 2014; Bauer et al. 2015; Chardin et al. 2017; Keating et al. 2018; Kulkarni et al. 2018) as they provide a cheaper alternative and a coarser resolution can be used to solve the radiative transfer, however they do not capture the full hydrodynamical evolution. For this reason, seminumerical and analytical methods of reionization modelling continue to remain attractive for efficient and flexible exploration of the parameter space (e.g. Furlanetto et al. 2004; Mesinger et al. 2011; Alvarez et al. 2012; Kaurov 2016; Kim et al. 2016; Hassan et al. 2016).

The main methodology, used in most hydrodynamical codes that aim for moderate to high spatial resolution, is to consider that all the gas in the simulation is optically thin to ionizing photons. This allows use of a uniform and isotropic UV+X-ray background radiation field (e.g. Katz et al. 1996) to define the ionization state of all gas elements. In this manner, all relevant information about the radiation field is enclosed by a set of photoionization and photoheating rates that evolve with redshift for each relevant ion (Haardt & Madau 1996, 2001; Faucher-Giguère et al. 2009; Haardt & Madau 2012). However, these UVB synthesis models surely break down during reionization events, and therefore the inhomogeneous nature of these processes is not captured (see e.g. the discussions in Oñorbe et al. 2017a; Puchwein et al. 2018). Currently, a common approach in reionization studies is to combine in post-processing the results of these types of hydrodynamical simulations, for the high-resolution description of the density fields, with seminumerical inhomogeneous models to study the various inhomogeneous effects (e.g Mesinger et al. 2015; Choudhury et al. 2015; Kulkarni et al. 2016)

The necessary next step to improve the predictions of numerical models is therefore to drop the assumption of a uniform and isotropic background in self-consistent hydrodynamical simulations. Motivated by this, we present in this paper a hybrid scheme for implementing inhomogeneous reionization physics into large-scale cosmological hydrodynamical simulations based on a small set of phenomenological input parameters. These models combine a seminumerical reionization model to solve for the morphology of reionization (Mesinger et al. 2011) with an approximate model of how reionization heats the IGM, using the massively parallel `Nyx` hydrodynamics code (Almgren et al. 2013), specifically designed to solve for the structure of diffuse IGM gas.

The structure of this paper is as follows. In Section 2 we describe our new method to model reionization in hydrodynamical simulations and compare it with other methods used in the literature. Section 3 describes the characteristics of all the hydrody-namical simulations studied in this work. Section 4 presents the results of simulations using one inhomogeneous and instantaneous reionization models but different heat inputs during reionization. We compare the outcome of different inhomogeneous reionization models in Section 5. In Section 6 we further investigate the differences between our simulations studying the properties of their temperature and density fields. Section 7 shows the effects that fluctuations of the UV background could have on the Ly$\alpha$ forest in both our flash and inhomogeneous reionization models. We discuss in Section 8 the limitations of our new approach and compare it with previous work. We conclude by presenting a summary of our results in Section 9. Finally in Appendix A we present a set of tests in which we run different variations of the new reionization modelling method introduced in this paper.

## 2 MODELLING REIONIZATION IN HYDRODYNAMICAL SIMULATIONS

In this paper we will focus on non-radiative transfer methods to implement reionization and the ultraviolet background (UVB) in hydrodynamical simulations that are suited to creating a large grid of different models and/or are easily used in combination with expensive subgrid star formation and feedback prescriptions. In this context we define three distinct approaches depending on how reionization is modelled in the simulations: *homogeneous* reionization, *inhomogeneous* reionization, and *flash* reionization.

### 2.1 Homogeneous reionization

A standard approach to modelling reionization and the UV background in hydrodynamical simulations is to assume that all resolution elements are optically thin to radiation. Thus, radiative feedback is accounted for via a spatially uniform, time-varying UVB radiation field, input to the code as a list of photoionization and photoheating rates that vary with redshift (e.g. Katz et al. 1996). These rates are obtained from 1D cosmological radiative transfer calculations (e.g. Faucher-Giguère et al. 2009; Haardt & Madau 2012; Khaire & Srianand 2018; Puchwein et al. 2018), also known as UVB synthesis models, and give the evolution of both the photoionization, $\Gamma_\gamma$ and the photoheating $\dot{q}$ rates with redshift for, at least, H I, He I and He II. This method is used in several hydrodynamical simulations that study galaxy formation and evolution (e.g. Schaye et al. 2015; Pillepich et al. 2018; Hopkins et al. 2018) and the circumgalactic medium (CGM) or the IGM (e.g. Lukić et al. 2015; Bolton et al. 2017; Oñorbe et al. 2017b). However, these UVB synthesis models surely break down during inhomogeneous reionization events and therefore their use in optically thin simulations has some intrinsic limitations. One of the most important issues is that the heat injection in this approach does not behave like realistic ionization fronts and it is done too gradually so it is therefore usually underestimated (see e.g. the discussion in Oñorbe et al. 2017a; Puchwein et al. 2018).

### 2.2 Inhomogeneous reionization

We introduce here a new method to account for inhomogeneous reionization in cosmological hydrodynamical simulations. A similar but simpler approach was introduced by Feng et al. (2016) based on seminumerical work by Battaglia et al. (2013). This method assigns a specific reionization redshift to each resolution element of the simulation and works as follows. When the reionization redshift of a resolution element is smaller than the current redshift of the simulation, the UVB seen by the resolution element is assumed to be zero. During the first time-step of the simulation in which the reionization redshift is crossed we inject a fixed amount of energy to account for the heating produced by the ionization front (Abel & Haehnelt 1999). We will first describe how we have implemented this heating in the simulation, and then discuss how we have generated the different reionization models.

The heating due to reionization is a complicated process that depends not only on the shape of the ionizing spectrum but also on the local density field and how fast the ionization front travels (Miralda-Escudé & Rees 1994; Abel & Haehnelt 1999; McQuinn 2012; Davies et al. 2016; D'Aloisio et al. 2018a). These studies have suggested a post-reionization temperature of order $\Delta T \sim 2 \times 10^4$ K, depending on the ionizing photon spectrum and the velocity of the ionization front[1]. In this work we choose as a first step to simply parametrize our ignorance of the details of reionization using heat injection as a free parameter, $\Delta T$. The temperature of the resolution element after reionization, $T^{post-reion}$, is then calculated as follows:

$$T^{post-reion} = (x_{H\,I}^{pre-reion})(\Delta T - T^{pre-reion}) + T^{pre-reion} \quad (1)$$

where $T^{pre-reion}$ and $x_{H\,I}^{pre-reion}$ are the temperature and the hydrogen ionization state just before reionization and $\Delta T$ is again a free parameter in our code. In this way, $\Delta T$ represents the maximum temperature that can be reached after reionization[2]. If the resolution element was already ionized the heating will be attenuated proportionally. This avoids any heat injection to regions that may have been already ionized by other physical processes included in the simulations, for example collisional ionization. Finally, we want to emphasize that we inject the necessary energy to ensure that the final temperature of the resolution element will be $T^{post-reion}$ taking into account its final ionization state, $x_{H\,I}^{post-reion}$, with the UVB already turned on.

While the density dependence of the heat injection during reionization may be the easiest further improvement that could be added to the model (see e.g. Kaurov 2016; Hirata 2018), its exact solution still depends on the assumed velocity of the ionization front. Therefore we have decided that in order to start exploring the parameter space, the best approach is to use the simplest model and assume that the reionization temperature is independent of density, which is common in studies of the thermal history of the IGM. We will discuss in detail the implications of these results for our methodology in Section 8.

After its reionization redshift, the resolution element is assumed to be optically thin and is exposed to the (uniform) tabulated photoionization rates given to the code. However, in contrast to the full homogeneous approach typically used in cosmological hydrodynamical simulations, in this method the UVB is only applied to the resolution elements that have already been reionized. Therefore, we need to provide the evolution of the average UVB in ionized regions and not just the evolution of the average UVB in the Universe. Of course, once reionization is finished both quantities will be the same but during reionization they will be very different (see e.g.

---
[1] Results from 3D radiative transfer cosmological simulations are still not that homogeneous in this regard. The treatment of multifrequency radiative transfer and particularly of the hard tail of the photon spectrum varies significantly among the methods and yields a correspondingly large range of temperatures after the reionization front (see e.g. Iliev et al. 2009).

[2] We discuss in Appendix A variations of this approach.

Lidz et al. 2007). For this reason we have used the photoionization and photoheating rates of Haardt & Madau (2012)[3] for $z \leq 6.0$. At higher redshifts, $z > 6$, we have assumed that the photoionization and photoheating in ionized regions are the same as at $z = 6.0$, which guarantees that all ionized regions have a H I ionized fraction of $x_{\rm H\,I} \simeq 10^{-4}$. While this approach does not directly include UVB fluctuations, they can be modelled straightforwardly in post-processing. We will present the results of adding UVB fluctuations in this way to our simulations in Section 7

We now describe how we assign a reionization redshift to each resolution element. We use a seminumerical method based on the excursion set formalism, applied to the initial conditions of the simulation, to pre-compute a reionization redshift field that is read at the beginning of the simulation. We describe in detail in §2.2.1 how we have done this. Alternatively, the reionization redshift could be computed while the simulation is running using the halo mass function computed directly on the fly, or some more advanced metrics if star formation is also modelled in the code (see e.g. Poole et al. 2016). We are already exploring this possibility and will present these results in the near future. In any case, we want to emphasize that the heat injection method described above is independent of how the reionization redshift is assigned.

### 2.2.1 Using seminumerical methods to generate H I reionization fields

Our method relies on having a unique reionization redshift for each resolution element in the simulation. In this paper we use a seminumerical approach to generate the reionization redshift field based on the analytical method introduced by Furlanetto et al. (2004) to describe the morphology of reionization on large scales. In this model each point at position $r$ has assigned a 1D function, trajectory $\Theta_r(R)$, corresponding to the average matter density within spheres of radius $R$ centred on $r$. The physical motivation is that the function $\Theta_r(R)$ can provide an estimate for the fraction of matter collapsed into halos (Press & Schechter 1974). Then, assuming a rate of ionizing photon production in halos, it is possible to derive the total number of ionizing photons in the spherical regions centred at the point of interest. Therefore as the density field is evolved, an ionization history can be obtained.

This approach is therefore physically based on the excursion set framework that is widely used in structure formation theory (Bond et al. 1991). It has become very popular in recent years because of its high computational efficiency. Various improvements can be made to this model, which allow the inclusion of more sophisticated physics (e.g. Sobacchi & Mesinger 2014). We have used the publicly available code `21cmFAST` (Mesinger et al. 2011) which is fundamentally based on the idea described above. In this code, the fraction of material in collapsed objects is computed analytically from an initial density field, which in our case corresponds to a particular realization of the high-$z$ initial conditions of the hydrodynamical simulation. To generate the reionization models used in this paper we have just freely varied two parameters: the minimum halo mass for producing ionizing photons, $M_{\rm min}$, and the ionizing efficiency, $\zeta$. We adopt the central-pixel-flagging algorithm instead of the slower bubble-painting algorithm (see Zahn et al. 2011, for a discussion of the relative accuracy of these methods) and a k-space top-hat filter when smoothing the density field. In this work we initialize `21cmFAST` using the same grid as in the `Nyx` simulations, $2048^3$, but generate evolved density and ionization fields on a coarser grid of $256^3$ (models A, B, D) or $128^3$ (model C). We generate ionization fields using temporal resolution of $\Delta z = 0.05$, going from $z = 15.0$ to $6.0$. From the combination of these ionization fields we construct a reionization redshift field that can be plugged in to the hydrodynamical simulation.

The left panel of Figure 1 shows a random slice of a reionization redshift field generated using the method described above. To generate this model we have used the $z = 159$ total density field of a $\Lambda$CDM cosmological box of size $L_{\rm box} = 40$ Mpc h$^{-1}$, $M_{\rm min} = 10^8$ $M_\odot$ and $\zeta = 11.0$. The exact values of these parameters are not relevant here and we will discuss them in detail below (see Section 3). This model produces a reionization history consistent with current CMB constraints (Planck Collaboration et al. 2018). The middle panel presents the same slice but now showing the total density field at $z = 6$ obtained running a cosmological hydrodynamical simulation. The main cosmological structures in the slice can be identified by eye, and it can be seen that as expected there is a correlation between dense structures and early values of the reionization redshift seen in the left panel. However, the right panel shows a Gaussian-smoothed version of the original density field at $z = 6$ with a filter of 1 Mpc h$^{-1}$. In this case, the correlation between the density values and the reionization map is even clearer.

The left panel of Figure 2 demonstrates that the relationship between density and reionization redshift is not simple. It displays the 2D histogram of the reionization redshift and the $z = 6$ density field. It is clear that there is no strong correlation between the reionization redshift of one resolution element and its precise density value at $z = 6$ (or at any other redshift). The middle and right panels of Figure 2 show the same slice again, but this time for a Gaussian-smoothed version of the total density field at $z = 6$ with a filter of 1 Mpc h$^{-1}$ and 5 Mpc h$^{-1}$, respectively. In this case there is a clear, although noisy, correlation between the reionization redshift of a resolution element and its specific smoothed density value. A similar correlation is found regarding of the redshift at which we pick the density field. This shows that the reionization redshift of each resolution element is correlated with the density field smoothed on large scales, as expected for reionization fields generated using the excursion set formalism (see e.g. Battaglia et al. 2013).

In this work we have created four different H I inhomogeneous reionization models using `21CMFAST` for various minimum halo masses: $M_{\rm min} = 10^8$, $10^6$ and $3 \times 10^9$ $M_\odot$ (models A, B and C respectively). We adjusted the ionization efficiency in these models so that all of them had the median reionization redshift as close as possible to $z_{\rm reion,H\,I}^{\rm median,inhomo} = 7.75$ while still having reionization finish at $z \simeq 6$ consistent with the Ly$\alpha$ forest. We also generated an extra model with $M_{\rm min} = 10^8$ with a faster reionization (model D, cyan solid line) using an ionizing efficiency evolving with redshift, $\zeta = 11.0 \times ([1+z]/[1+7.75])^{-4}$. The upper panel of Figure 3 shows the ionization fraction evolution for each of these models (green, pink and orange solid lines). The lower panel shows the evolution of the CMB electron scattering optical depth[4], $\tau_e$, for each of these models. The gray band shows the most recent constraints on $\tau_e$ from CMB data (Planck Collaboration et al. 2018). Table 1 contains a list of various parameters characterizing these reionization models: the median reionization redshift, $z_{\rm reion,H\,I}^{\rm median,inhomo}$, the width of H I reionization, $\Delta z_{\rm reion, H\,I} = z^{\langle x_{\rm H\,II}\rangle=0.1} - z^{\langle x_{\rm H\,II}\rangle=0.99}$, the Thompson scattering optical depth, $\tau_e$, and the end of H I reionization redshift, $z_{\rm reion, H\,I}^{\rm end}$.

---

[3] With the minor corrections suggested by (Oñorbe et al. 2017b) to provide a better fit to the observed H I and He II mean flux evolution.

[4] We used the results of the hydrodynamical simulations to compute the density-weighted ionization history for this model, which is the relevant ionization history for computing $\tau_e$ (see e.g. Liu et al. 2016).

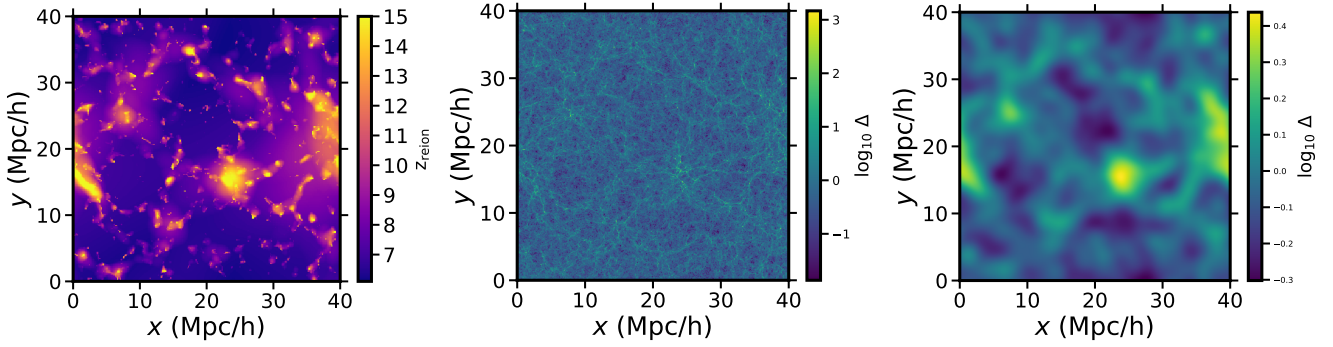

**Figure 1.** Left panel: a $\Delta z = 20$ kpc h$^{-1}$ slice of the reionization redshift field generated using the method described in Section 2.2.1 (model A). The evolution of the H I ionization fraction for this model can be found in Figure 3. The exact parameters used to generate this field are summarized in Table 1. Middle panel: the same slice as in the left panel, now showing the total density field at $z = 6$. Right panel: the same slice as presented in the left panel this time showing the total density field at $z = 6$ to which we have applied a Gaussian-smoothed filter of 1 Mpc h$^{-1}$.

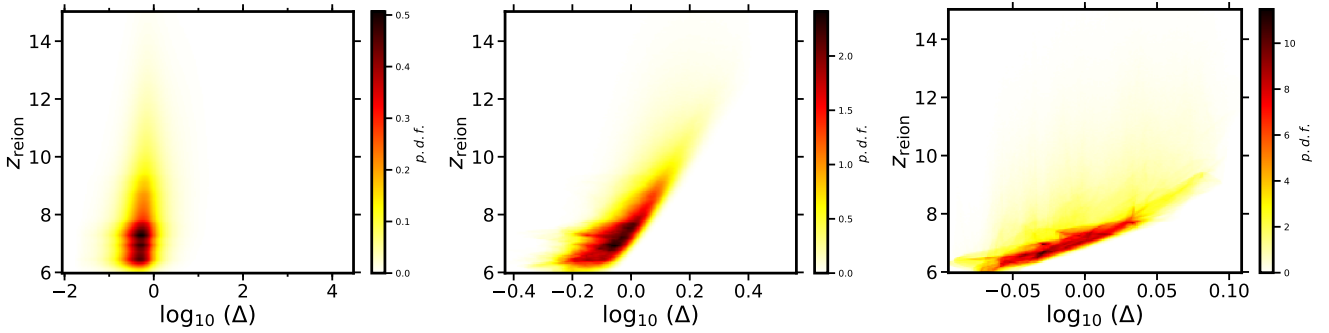

**Figure 2.** Left panel: the 2D histogram of model A reionization redshift and the $z = 6$ density fields of a cosmological hydrodynamical simulation using this model. No strong correlation between the reionization redshift of one resolution element and its precise density value is found. Middle and right panels: 2D histogram of the model A reionization redshift and the $z = 6$ smoothed density field with a gaussian smoothed filter of 1 Mpc h$^{-1}$ (middle) and 5 Mpc h$^{-1}$ (right). This shows that the reionization redshift of each resolution element is correlated with the density field smoothed on large scales, as expected for reionization fields generated using the excursion set formalism. See text for more details.

### 2.3 Flash Reionization

We consider one more method, "flash", to simulate the reionization process in hydrodynamical simulations. It relies on the same methodology to inject heat, but has a fixed reionization redshift for all resolution elements in the simulation. Therefore these simulations will have only two parameters to describe the reionization process: the redshift of reionization, $z_{\rm reion}^{\rm flash}$, and the heat injection, $\Delta T$.

Of course this method misses by construction the inhomogeneities associated with H I reionization. However, it allows us to easily compare them with the inhomogeneous approach and to study in more detail the effect that these inhomogeneities can have on different scales. Figure 3 shows a flash H I reionization evolution model that happens at $z_{\rm reion, H\,I}^{\rm flash} = 7.75$ (black line). Note that this value is very close to the median reionization redshift of the four inhomogeneous runs, $z_{\rm reion, H\,I}^{\rm median, inhomo}$.

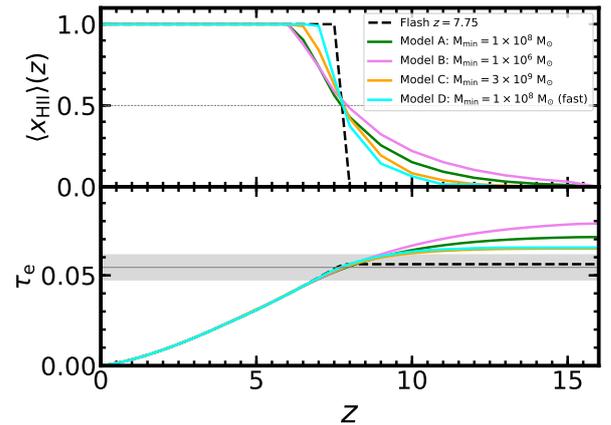

**Figure 3.** Reionization models studied in this work. Top panel: evolution of the H I ionization fraction for the various reionization models. Note that the models were built to have a similar median reionization redshift, $z_{\rm reion, H\,I}^{\rm median, inhomo} \simeq 7.75$. Lower panel: the integrated electron scattering optical depth, $\tau_e$, calculated from the above models using the density-weighted ionization fraction in the simulation. The gray band denotes the latest constraints on $\tau_e$ from Planck Collaboration et al. (2018) data.

**Table 1.** Summary of simulations.

| Sim | Method | $z^{\text{median}}_{\text{reion,H\,I}}$ | $z^{\text{end}}_{\text{reion,H\,I}}$ | $\Delta z_{\text{reion,H\,I}}$ | $\tau_e$ | $\Delta T_{\text{H\,I}}$ [K] | $\zeta$ | $M_{\text{min}}$ [$M_\odot$] |
|---|---|---|---|---|---|---|---|---|
| IR-A | Inhomogeneous (Model A) | 7.75 | 6.05 | 4.82 | 0.7120 (0.06260) | $2 \times 10^4$ | 11.0 | $1 \times 10^8$ |
| IR-Acold | Inhomogeneous (Model A) | 7.75 | 6.05 | 4.82 | 0.7120 (0.06260) | $\mathbf{1 \times 10^4}$ | 11.0 | $1 \times 10^8$ |
| IR-Ahot | Inhomogeneous (Model A) | 7.75 | 6.05 | 4.82 | 0.7120 (0.06260) | $\mathbf{4 \times 10^4}$ | 11.0 | $1 \times 10^8$ |
| IR-B | Inhomogeneous (**Model B**) | 7.90 | 6.05 | 6.06 | 0.07864 (0.06705) | $2 \times 10^4$ | 5.0 | $\mathbf{1 \times 10^6}$ |
| IR-C | Inhomogeneous (**Model C**) | 7.85 | 6.25 | 3.44 | 0.06488 (0.06067) | $2 \times 10^4$ | 45.0 | $\mathbf{3 \times 10^9}$ |
| IR-D | Inhomogeneous (**Model D**) | 7.63 | 6.85 | 2.55 | 0.06540 (0.05998) | $2 \times 10^4$ | evol. | $1 \times 10^8$ |
| FR | Flash | 7.75 | 7.75 | 0.0 | 0.05670 | $2 \times 10^4$ | – | – |
| FRcold | Flash | 7.75 | 7.75 | 0.0 | 0.05670 | $\mathbf{1 \times 10^4}$ | – | – |
| FRhot | Flash | 7.75 | 7.75 | 0.0 | 0.05670 | $\mathbf{4 \times 10^4}$ | – | – |

Column 1: simulation code.
Column 2: method used to simulate reionization (see text for details).
Column 3: H I reionization median redshift.
Column 4: end of H I reionization.
Column 5: width of H I reionization, $\Delta z_{\text{reion,H\,I}} = z^{\langle x_{\text{H\,II}}\rangle=0.1} - z^{\langle x_{\text{H\,II}}\rangle=0.99}$.
Column 6: Thompson scattering optical depth $\tau_e$ density-weighted (volume-weighted).
Column 7: total heating assumed for H I reionization.
Column 8: parameter excursion set model 1, ionizing efficiency.
Column 9: parameter excursion set model 2, minimum mass.
All simulations have a box size of length, $L_{\text{box}} = 40$ Mpc h$^{-1}$ and $2048^3$ resolution elements.

## 3 SIMULATIONS

The simulations in this work were run with the Nyx code, a massively parallel N-body gravity + grid hydrodynamics code specifically designed for simulating the Ly$\alpha$ forest (Almgren et al. 2013). Nyx also includes the chemistry of the primordial composition gas, and inverse Compton cooling off the microwave background and keeps track of the net loss of thermal energy resulting from atomic collisional processes. We used the updated recombination, collision ionization, dielectric recombination rates and cooling rates given in Lukić et al. (2015). For more details of these numerical methods and scaling behavior tests, see Almgren et al. (2013).

We generated the initial conditions using the MUSIC code (Hahn & Abel 2011). We ran the CAMB code (Lewis et al. 2000; Howlett et al. 2012) to create the transfer function. The initial redshift for all simulations was $z_{\text{ini}} = 159$. In this work we used a standard $\Lambda$CDM cosmological model consistent, within one sigma, with the latest cosmological constraints from the CMB (Planck Collaboration et al. 2018): $\Omega_m = 0.3192$, $\Omega_\Lambda = 0.6808$, $\Omega_b = 0.04964$, $h = 0.67038$, $\sigma_8 = 0.826$ and $n_s = 0.9655$. We adopted the following hydrogen and helium mass abundances: $X_p = 0.76$ and $Y_p = 0.24$ respectively (therefore $\chi = 0.0789$), to match recent CMB observations and Big Bang nucleosynthesis (Coc et al. 2013). All simulations were run down to $z = 4.0$, saving 32 snapshots[5]. Unless otherwise stated, all simulations presented in this work have a box size of length, $L_{\text{box}} = 40$ Mpc h$^{-1}$ and $2048^3$ resolution elements.

We ran hydrodynamical simulations with the four inhomogeneous reionization models (IR-A, IR-B, IR-C, IR-D, named after the reionization models used in them, see Figure 3), and the flash reionization model (FR) described in Section 2.2. In all these runs we assumed a total heat injection of $\Delta T_{\text{H\,I}} = 2 \times 10^4$ K, which is the standard value obtained in galaxy-driven H I reionization models (see above). Quasar- or PopulationIII-driven scenarios may inject more heat, as much as $\Delta T_{\text{H\,I}} \sim 4 \times 10^4$ K (but see D'Aloisio et al. 2018a, and the discussion in Section 8). So in order to study the effects of different heat injections during H I reionization we also ran the flash reionization model, $z^{\text{flash}}_{\text{reion,H\,I}} = 7.75$, and model A inhomogeneous reionization ($M_{\text{min}} = 10^8 \, M_\odot$) with two different values for the heat injected during reionization: a colder model with $\Delta T_{\text{H\,I}} = 1 \times 10^4$ K and a hotter model with $\Delta T_{\text{H\,I}} = 4 \times 10^4$ K (hereafter referred as IR-Acold and IR-Ahot for the inhomogeneous runs and as FR-cold and FR-hot for the flash runs respectively).

## 4 COMPARISON OF FLASH AND INHOMOGENEOUS REIONIZATION SIMULATIONS

We now present the results of the inhomogeneous and flash reionization simulations. We will first focus on the differences between these two methods. Figure 4 shows the same slice of the temperature field of our fiducial flash reionization simulation (FR, left column) and inhomogeneous reionization simulation (IR-A, right column) at $z = 7$, $z = 6$, $z = 5$ and $z = 4$ (from top to bottom respectively). These two simulations have the same heat injection, $\Delta T$, and the H I reionization redshift of the flash simulations is equal to the median reionization redshift of the inhomogeneous reionization, $z^{\text{flash}}_{\text{reion,H\,I}} = z^{\text{median,inhomo}}_{\text{reion,H\,I}} = 7.75$. Therefore at $z = 7$ reionization has just finished in the flash reionization and we can see an overall homogeneous high temperature (top left panel), while at the same redshift the inhomogeneous reionization (top right panel) is still ongoing and we can clearly see the difference in temperatures between regions that have been reionized and those that remain neutral. At $z = 5$ the temperature field in the inhomogeneous approach (third row left column panel) shows lingering signatures of the temperature fluctuations on $\sim 5 \, h^{-1}$ Mpc scales. In this run different regions of the Universe are reionized and heated at different times, and so they asymptote to the temperature set by the balance between photoheating and the adiabatic and Compton cooling governing the temperature-density relationship at different times (McQuinn & Upton Sanderbeck 2016). In this particular case we see that the high-density regions, which reionized at high redshift have had time to cool and show lower temperatures than the low-density regions, which were reionized much later (D'Aloisio et al. 2015; Davies et al. 2018a). This will have implications for the properties of Ly$\alpha$ at these scales, as will be discussed in detail in Section 4.1. The high-temperature filamentary structure observed in all panels of Figure 4 reflects the high-density regions where collapsed objects have heated the gas to their virial temperatures. This is of course independent of the reionization method employed[6].

---
[5] We stored snapshots at the following redshifts: 20, 19, 18, 17, 16, 15, 14, 13, 12, 11, 10, 9, 8, 7.5, 7, 6.5, 6, 5.8, 5.6, 5.4, 5.2, 5, 4.8, 4.6, 4.4, 4.2, 4.

It is also interesting to study the differences in the temperature-density relationship between these two runs. Figure 5 shows the volume-weighted temperature-density histogram for the same flash (FR, left column) and inhomogeneous (IR-A, right column) simulations at $z = 7$, $z = 6$, $z = 5$ and $z = 4$ (from top to bottom, respectively). At $z = 7$, reionization has just happened in the flash simulations ($z_{\rm reion,H\,I}^{\rm flash} = 7.75$) and therefore most of the gas is still at high temperatures, leading to a flat temperature-density relationship. As time evolves, the adiabatic and Compton cooling is efficient enough to give rise to a steeper temperature-density relationship despite the photoheating background. For the inhomogeneous run, however, the temperature-density relationship is not well defined until reionization is finished ($z = 6.05$ in this model). Before this time, the temperature-density distribution is bimodal, reflecting regions that have experienced reionization versus the regions that still have not. This is clearly seen in the $z = 7$ temperature-density distribution (lower left panel of Figure 5). Even when reionization is finished the inhomogeneous simulation shows a much larger scatter than the flash simulation, especially at lower densities (see Trac et al. 2008; Lidz & Malloy 2014; Keating et al. 2018, for similar findings). This is produced because, as shown in Section 2, for a fixed density there is a wide range of reionization redshifts and we can see the effect of the different timing of the heating and cooling process for all resolution elements. Towards lower redshift the scatter is reduced, but even at $z = 4$ we still see a significant difference between the inhomogeneous run and the flash run (see also Trac et al. 2008, who found similar effects using radiative transfer simulations).

We have also measured at each snapshot the volume-weighted temperature at mean density, $T_0$, and the slope of the power-law temperature-density relationship, $\gamma$, where $T = T_0 \Delta^{\gamma-1}$. The parameters are obtained by fitting the temperature-density relationship with linear least squares in $\log_{10} \Delta$ and $\log_{10} T$ using only cells that satisfy the following criteria: $-0.7 < \log_{10} \Delta < 0.0$ and $\log_{10} T/K < 4.5$. The panels in Figure 5 show as a black solid line the results for both simulations. Changing these thresholds for the flash simulation within reasonable IGM densities ($\log \Delta < 0.7$) and temperatures ($\log_{10} T/K < 7$) produces differences just at the few per cent level at any redshift after reionization. We have tested that we obtain similar conclusions if we change the fitting method and obtain the thermal parameters, finding the median of the gas temperature in bins of width 0.02 dex at $\log \Delta = 0$ and 0.7. We find no relevant changes if we modify these two densities slightly (again as long as we do not include too high densities, $\log \Delta < 0.7$). We find, however, that the fitting procedure used seems to have a more important role in the value of the thermal parameters in the inhomogeneous runs. This is true not only before reionization is complete but also at later times. At $z = 7.0$ our inhomogeneous reionization is not yet complete, so it could be argued that in order to define a clear temperature-density relationship only ionized gas should be used (top right panel of Figure 5). Once reionization is complete, we find that different fitting methods and parameters result in $\lesssim 10\%$ differences of $T_0$ and $\gamma$ at $z = 6$ and in $\lesssim 5\%$ differences at $z = 4$. Note, however, that these differences are systematics for all simulations. We find that the parameters with the largest effect on the temperature-density relationship fit are the density thresholds. Although this effect has no relevant consequences for the results and conclusions presented in this work it will have to be considered in future studies of the IGM thermal parameters that use simulations of inhomogeneous reionization.

In Figure 6 we present the thermal histories for a set of flash (dashed lines) and inhomogeneous (solid lines) reionization simulations. All flash simulations share the same reionization redshift $z_{\rm reion,H\,I}^{\rm flash} = 7.75$ but they differ in the heat injected during reionization: $\Delta T = 10^4$ K, $\Delta T = 2 \times 10^4$ K, $\Delta T = 4 \times 10^4$ K (blue, green and red, respectively). We show the analogous inhomogeneous reionization simulations with different heat injection during reionization and the same reionization evolution in which $z_{\rm reion,H\,I}^{\rm median,inhomo} = z_{\rm reion,H\,I}^{\rm flash} = 7.75$. For the inhomogeneous runs we only show the evolution once reionization is finished ($z < 6.05$) as before that time such global parameters are not very meaningful. The top and middle panels of Figure 6 show the evolution of the $T_0$ and $\gamma$ thermal parameters (respectively) governing the temperature-density relationship determined by fitting the distribution of densities and temperatures in the simulation discussed above. As a reference we also show in the top panel the median values of temperature (black dots) and the 1$\sigma$ scatter (black solid errorbars) that we found for the IR-A inhomogeneous model at $z = 4$, $z = 5$ and $z = 6$.

The bottom panel of Figure 6 shows the evolution of the pressure smoothing scale, $\lambda_{\rm P}$, as defined by Kulkarni et al. (2015). This parameter is defined using a pseudo real-space Ly$\alpha$ flux field, obtained assuming no redshift space effects. This allows the effect of the dense gas dominating the baryon power spectrum to be cancelled, which reveals the pressure smoothing in the diffuse IGM.

Figure 6 shows that, regardless of the reionization method implemented in the simulation, models with less heat injection have lower temperatures and slightly higher $\gamma$ values. These models also produce lower pressure smoothing scales, but in this case the differences continue after the other thermal parameters, $\gamma$ and $T_0$, have converged. All these models have identical photoionization and photoheating rates when reionization ends and therefore $\gamma$ and $T_0$ converge to the same values faster than the pressure smoothing scale, which has memory of the thermal history (Kulkarni et al. 2015; Nasir et al. 2016; Oñorbe et al. 2017b). While $T_0$ is very similar between the analogous flash and inhomogeneous models, this is not the case for $\gamma$ and inhomogeneous simulations show systematically lower values. This is expected from the study of the temperature-density relationship presented above (Figure 5) as we have seen that the slope of the temperature-density relationship is more sensitive to the larger scatter in this relationship that appears in inhomogeneous reionization models. However, the evolution of the pressure smoothing scale, $\lambda_{\rm P}$, unlike the other thermal parameters, is very different between the flash and inhomogeneous models that share the same amount of heat injection during reionization. In the next section we consider how the differences in thermal parameters between these simulations translate into Ly$\alpha$ statistics. We return to a more detailed discussion of the statistical properties of the temperature and density field in Section 6.

---

[6] Note that, in contrast to some other models such as for example Trac et al. (2008), our simulations do not include any kind of thermal feedback in the dense regions from star formation.

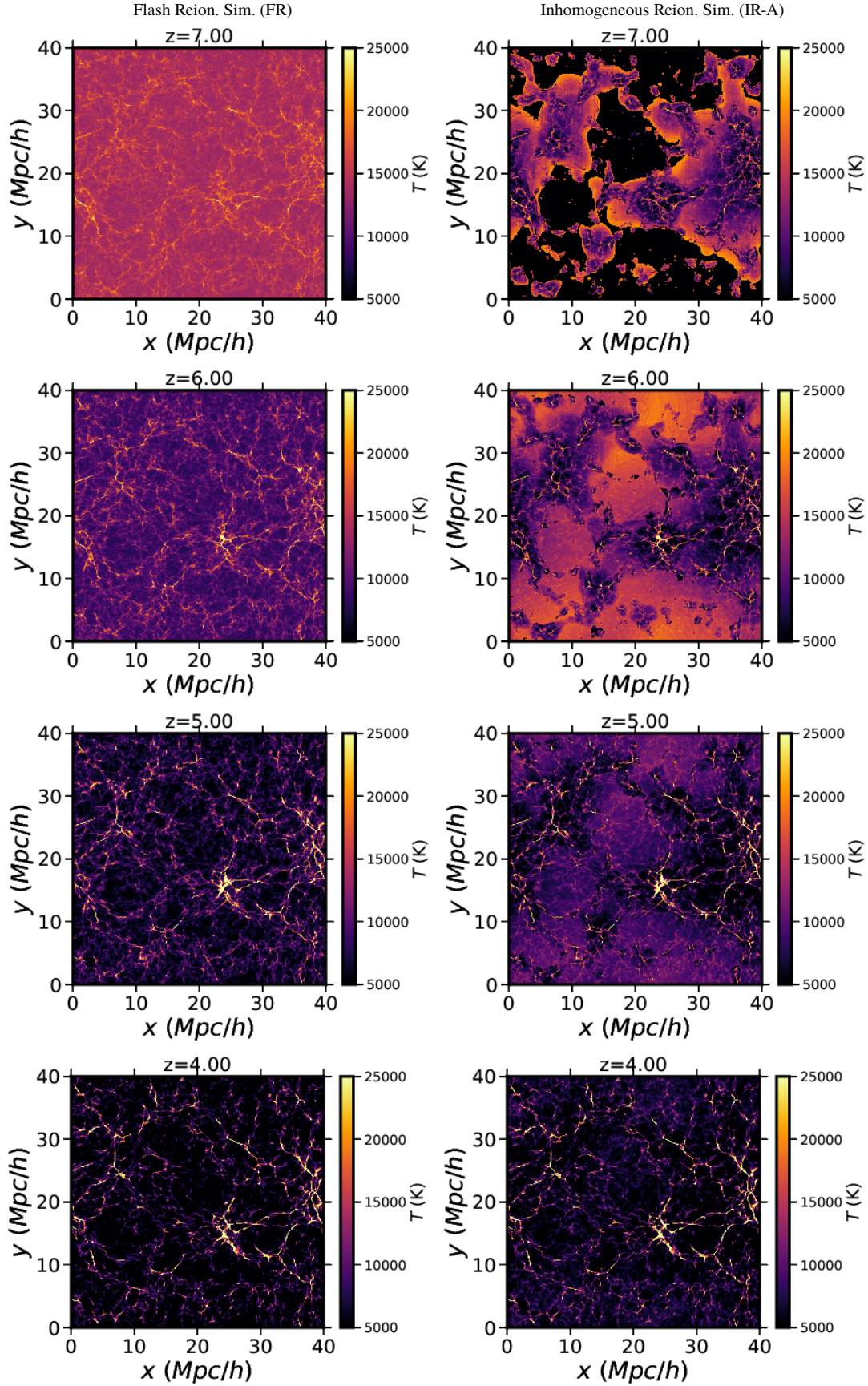

**Figure 4.** The same slice of the temperature field for our fiducial flash (FR, left column) and inhomogeneous (IR-A, right column) reionization simulations at redshifts $z = 7$, $z = 6$, $z = 5$ and $z = 4$ (from top to bottom respectively). Thermal fluctuations are clearly seen in the inhomogeneous run well after reionization is finished, with hotter (colder) regions tracing late (early) reionization regions.

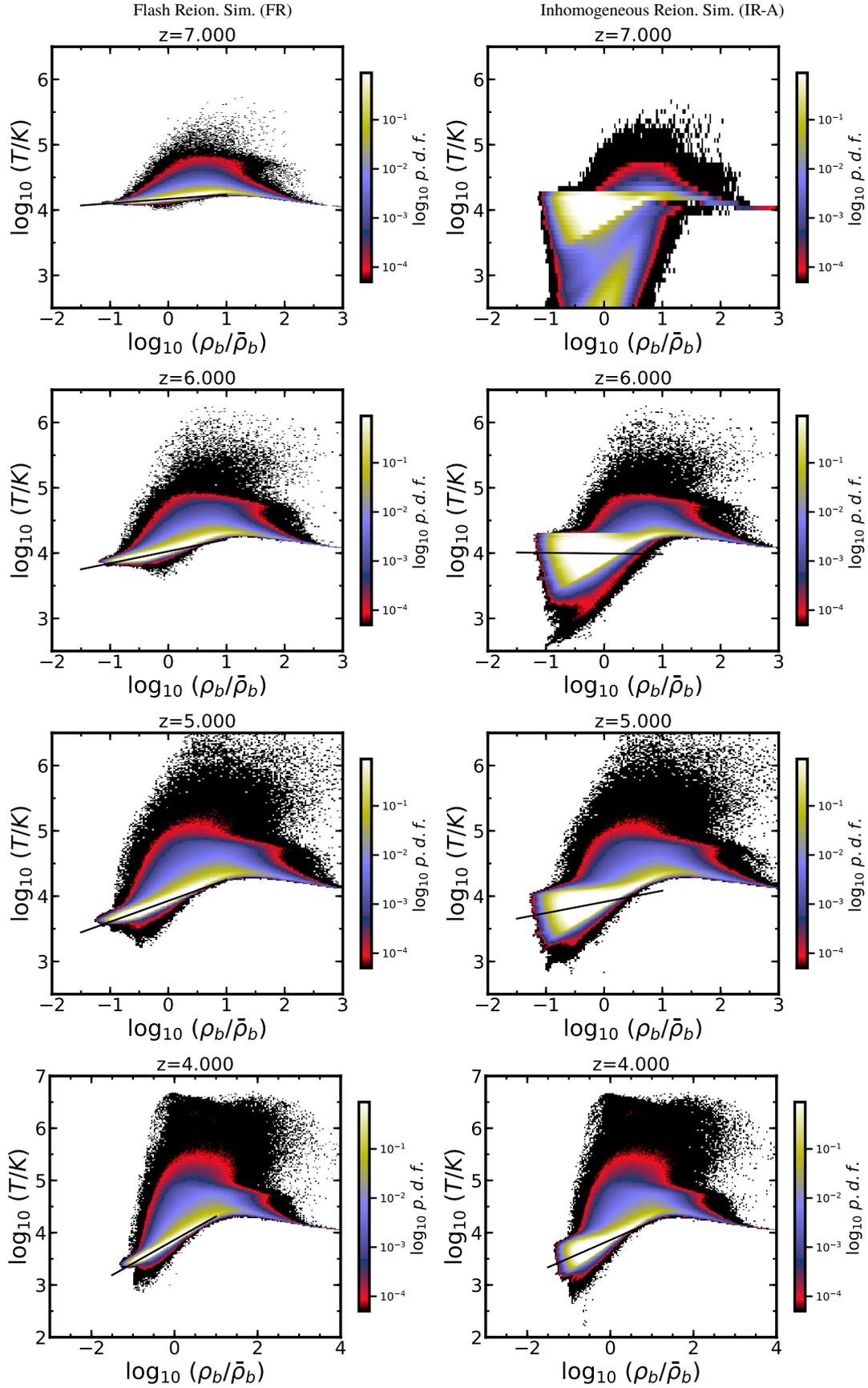

**Figure 5.** The temperature-density relationship for our fiducial flash (FR, left column) and inhomogeneous (IR-A, right column) reionization simulations at redshifts $z = 7$, $z = 6$, $z = 5$ and $z = 4$ (from top to bottom respectively). At $z = 7$ reionization has just happened in the flash run ($z_{\rm reion, H\,I}^{\rm end} = 7.75$) while the inhomogeneous run still shows a bimodal distribution. Even after reionization is finished in the inhomogeneous run ($z = 6.05$) a much wider temperature-density distribution is found in the inhomogeneous simulation.

### 4.1 Lyman-$\alpha$ forest statistics: the 1D power spectrum

In order to further explore the differences between the inhomogeneous and flash reionization simulations presented above, we now investigate the properties of their respective Ly$\alpha$ forests. To compute the Ly$\alpha$ forest spectra from the simulation, we compute the H I optical depth at a fixed redshift accounting for thermal broadening and Doppler shifts following Lukić et al. (2015). We can then convert into a transmitted flux fraction, $F = e^{-\tau}$ as a function of wavelength (or time or distance). As is standard procedure, we adjust the UVB so that the mean flux of the extracted spectra matches the observed mean flux at each redshift. We did not consider noise or metal contamination in our skewers, but this does not affect any of the main results of this work.

For each simulation we computed the 1D flux power spectrum, $P(k)$, of the fractional contrast, $\delta F$, at each redshift, defined as $\delta F = F/\langle F \rangle - 1$. We computed the power spectrum of $2048^2$ skewers at each redshift with a length equal to the size of the box and then calculated the average value at each mode, $k$. The 1D flux power spectrum is known to be sensitive to the thermal state of the IGM, because the Doppler broadening and the pressure smoothing reduce the amount of small-scale structure and generate a cut-off in the flux power spectrum $P(k)$ at small scales (high-$k$) (Zaldarriaga et al. 2001; Peeples et al. 2010; Walther et al. 2018a). For this reason, the detailed study of this cut-off can provide constraints not only on the thermal state of the gas but also on its full thermal history (Nasir et al. 2016; Oñorbe et al. 2017b; Boera et al. 2018). Moreover, the large-scale properties of Ly$\alpha$ forest (low-$k$) could be used to study possible fluctuations of the UVB or the temperature field that could provide clues about possible sources responsible for reionization (e.g. D'Aloisio et al. 2018b).

In this work we assumed the following mean flux values at each redshift to be consistent (within $1\sigma$) with recent observational constraints. This level of accuracy is more than enough for this work and none of the results presented in this paper depend on the exact values assumed for the mean flux. At $z = 4$, and $z = 4.2$ we used $\langle F \rangle(z = 4.0) = 0.411$ and $\langle F \rangle(z = 4.2) = 0.364$ respectively. This is in good agreement (within $1\sigma$) with observations (Becker & Bolton 2013; Viel et al. 2013a; Eilers et al. 2018). At $z = 5$ and $z = 5.4$ we used $\langle F \rangle(z = 5) = 0.14$ and $\langle F \rangle(z = 5.4) = 0.08$, respectively, consistent with recent measurements (D'Aloisio et al. 2018b; Bosman et al. 2018) and just above the $1\sigma$ limit of Eilers et al. (2018). Finally, for $z = 6$ we assumed a mean flux of $\langle F \rangle(z = 6) = 0.011$. This is slightly above the $1\sigma$ range measured by Bosman et al. (2018, $\langle F \rangle(z = 6) = 0.007^{+0.003}_{-0.002}$) and Eilers et al. (2018, $\langle F \rangle(z = 6) = 0.0052 \pm 0.0043$).

Figure 7 presents the dimensionless 1D flux power spectrum, $kP(k)/\pi$, computed at $z = 6.0$, $z = 5.0$, and $4.0$ (left, middle and right panels respectively) for the various flash (dashed lines; FR, FRcold, FRhot) and inhomogeneous (solid lines; IR-A, IR-Acold, IR-Ahot) simulations. The color of each line indicates the heat injected during reionization: $\Delta T = 10^4$ K (blue), $\Delta T = 2 \times 10^4$ K (green) and $\Delta T = 4 \times 10^4$ K (red). The first thing to notice between the three panels is that the overall power level increases with $z$, which is due to the average mean flux decreasing towards higher redshift, exponentially amplifying the density fluctuations (e.g. Viel et al. 2004; Palanque-Delabrouille et al. 2013; Viel et al. 2013b; Walther et al. 2018a). At each redshift, the lower panels show the percentage difference between each flash simulation and its analogous inhomogeneous run, that is, the one that has the same heat input during reionization, $\Delta T$. At low-$k$ modes (large scales) there is a systematic difference between the reionization approaches, and the inhomogeneous runs show more power at these scales than their flash counterparts. This is due to the large-scale coherence of temperature fluctuations, which can be linked to the specific details of the reionization model and the sources responsible for it. It can be seen that both the overall difference in power and the $k$ mode at which this effect is relevant decrease with redshift. This decrease shows how the temperature fluctuations are being attenuated in the simulation as time evolves. As expected, inhomogeneous models with a larger total heat injection during reionization show larger differences from the flash reionization model because the temperature fluctuations in these models are also larger. Note that the difference between the flash and inhomogeneous models gradually increases towards lower $k$ modes. At high-$z$, $z \gtrsim 5$, the difference is as large as $\sim 50\%$ for the lowest modes that we can study in these simulations ($k \sim 1 \times 10^{-3}$ s km$^{-1}$). We expect these differences to continue increasing as we go to lower $k$ modes (e.g. D'Aloisio et al. 2018b). However, note that the overall power decreases as we go to larger scales (lower-$k$), and Ly$\alpha$ forest observations are also limited on the minimum scale that can be measured accurately which is about $\sim 2 \times 10^{-3}$ s km$^{-1}$ (Lee et al. 2012; Palanque-Delabrouille et al. 2013; D'Aloisio et al. 2018b). We will compare current high-$z$ large-scale measurements with our models in the next section.

Regardless of the reionization methodology used, simulations with a lower heat input during reionization have the scale cut-off of the power spectrum at higher-$k$ modes (small scales). This can be explained because different heat injections lead to different amounts of pressure smoothing and thermal broadening (see Fig. 6) that change the shape of the cut-off in the power spectra (Oñorbe et al. 2017b). Interestingly, the power spectrum at $0.02 < k < 0.1$ s km$^{-1}$ is very similar for inhomogeneous and flash models when using the same heat input. At $z = 6$ the largest difference is $\sim 10\%$ between the models with the largest heat input during reionization and it decreases towards lower redshifts. We found that this similarity occurs when the inhomogeneous and flash models have the same heat input owing to reionization, $\Delta T$, and the median redshift of reionization of the inhomogeneous run is equal to the reionization redshift of the flash simulation; that is, $z_{\rm reion, H I}^{\rm median, inhomo} = z_{\rm reion, H I}^{\rm flash}$. This has very important implications because it suggests that the thermal fluctuations associated with the inhomogeneous reionization models do not significantly change the small-scale correlation properties of the Ly$\alpha$ forest. It shows that once reionization is finished the small scale properties of the Ly$\alpha$ forest in an inhomogeneous reionization model can be reasonably well described with a flash model that shares the same average heat injection and reionization time. It is likely that the lowest $k$ (i.e., the largest scale) at which both types of models agree depends on the exact morphology of the reionization redshift field and the relevant scale of temperature fluctuations that arises from it. Below we will further explore this interesting result using different inhomogeneous reionization models.

## 5 DIFFERENT INHOMOGENEOUS MODELS

We now turn to the results of our simulations using different inhomogeneous reionization models: IR-A, IR-B, IR-C and IR-D in which we have varied the dominant halo masses responsible for H I reionization ($M_{\rm min} = 10^8$, $10^6$ and $3 \times 10^9$ $M_\odot$ respectively) and adjusted the ionizing efficiency $\zeta$ such that reionization is half complete ($x_{\rm H I} = 0.5$) at a very similar redshift (i.e. all have a very similar $z_{\rm reion, H I}^{\rm median, inhomo}$, see Figure 3 and Section 2 for further details). We ran all these models with the same heat injection during reionization, our fiducial value $\Delta T = 2 \times 10^4$ K. To illustrate the differences between these runs, the upper row of Figure 8 shows the same random slice of the reionization redshift field for the four simulations. These slices reveal the different characteristic morphologies for these models, where we can see a clear trend with the characteristic minimum halo mass assumed to create them. As expected, larger minimum halo masses result in larger characteristic ionized bubbles. Figure 8 shows the same slices of the $x_{\rm H\,I}$ (middle row) and temperature (bottom row) fields for these runs at $z = 8$, when the ionization fraction between the models is very similar. In the lower row it can be seen that, as expected, the morphology in the $x_{\rm H\,I}$ field translates directly to the temperature field at these redshifts. The different heating times produce temperature fluctuations persisting to lower redshifts. We now explore whether we can use the high-$z$ Ly$\alpha$ forest to discriminate between these different reionization histories and morphologies. First we will check the evolution of the thermal parameters of these models.

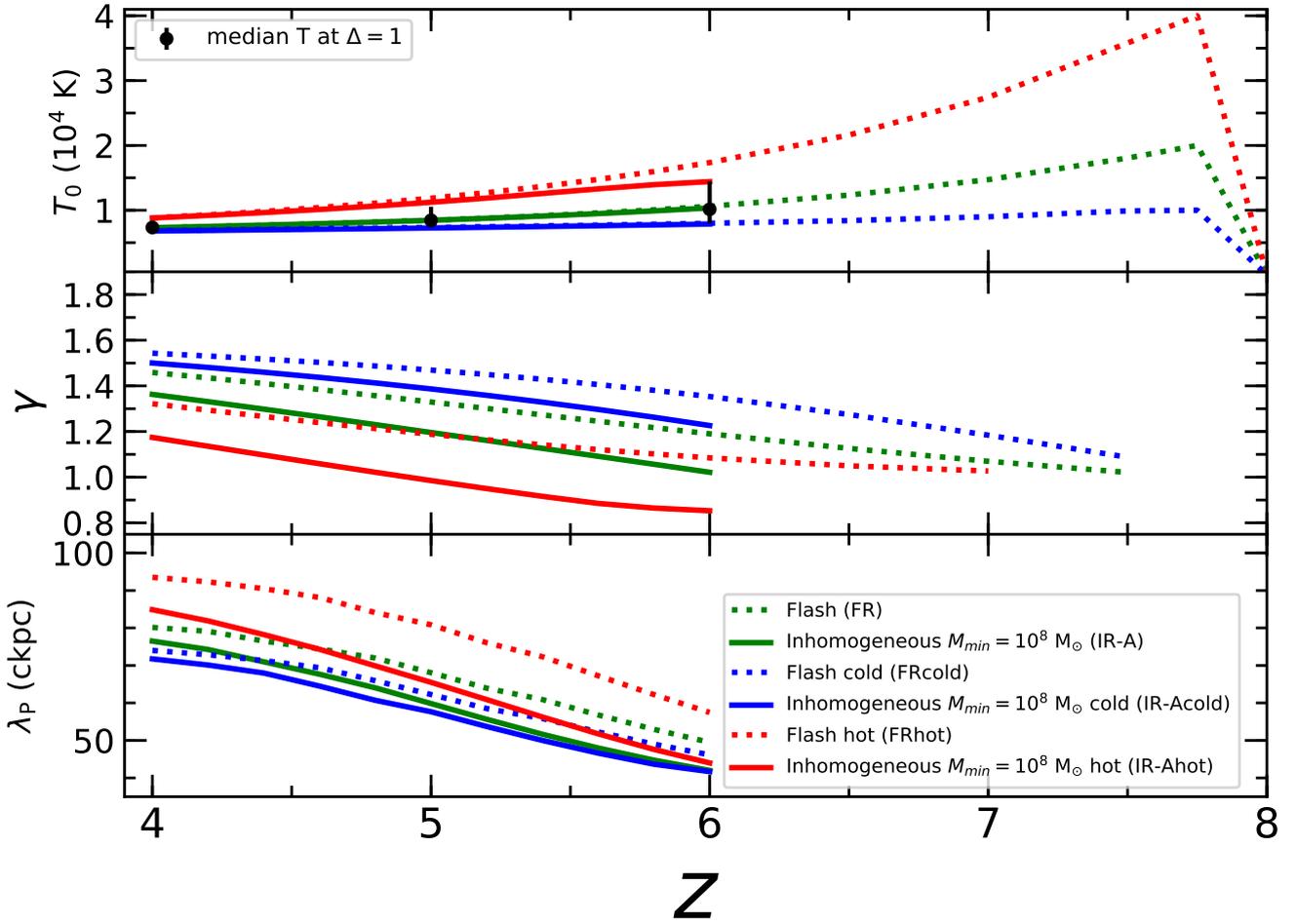

**Figure 6.** The evolution of thermal parameters with redshift for a set of flash (dashed lines) and inhomogeneous models (solid lines). Models with the same color have the same heat injection during reionization: $\Delta T = 10^4$ K (blue), $\Delta T = 2 \times 10^4$ K (green) and $\Delta T = 4 \times 10^4$ K (red). From top to bottom it shows the temperature at mean density, $T_0$, the slope of the temperature-density relationship, $\gamma$ and the pressure smoothing scale, $\lambda_P$, as defined in Kulkarni et al. (2015). In the first panel, the black circles and error bars show the median value of the temperature at mean density and the $1\sigma$ scatter found for the IR-A run at $z = 4$, $z = 5$ and $z = 6$. See text for more details.

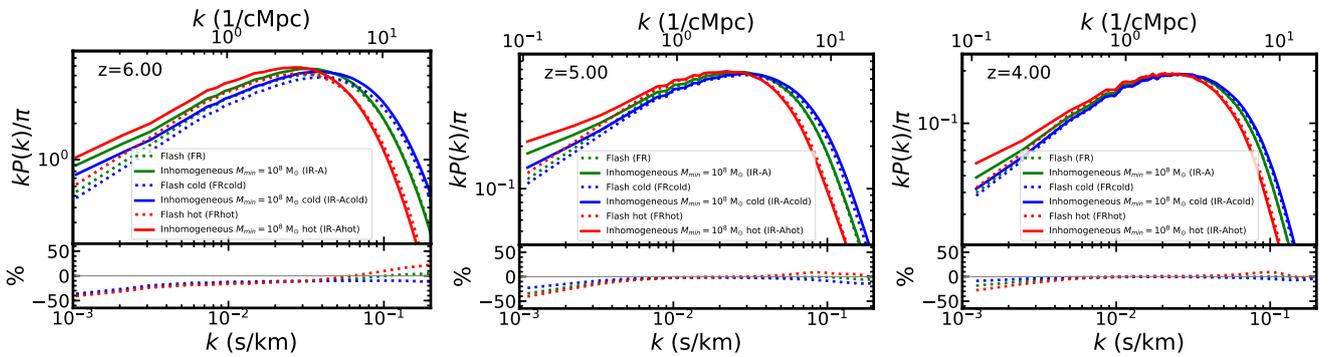

**Figure 7.** The 1D flux power spectrum at $z = 6.0$, $z = 5.0$ and $4.0$ (from left to right respectively) for a set of different flash (dashed lines) and inhomogeneous (solid lines) models. The color of each line denotes the heat injected during reionization: $\Delta T = 10^4$ K, $\Delta T = 2 \times 10^4$ K, $\Delta T = 4 \times 10^4$ K (blue, green, red respectively). The reionization redshift of the flash reionization simulations agrees with the median reionization redshift of the inhomogeneous runs. The lower sections of each panel show the percentage difference between each flash simulation and its analogous inhomogeneous run, namely the one with the same heat input during reionization. Note that inhomogeneous and flash reionization models differ at low-$k$ modes (large scales) owing to temperature fluctuations but still agree at high-$k$ modes (small scales).

Figure 9 shows the evolution of the thermal parameters for the various inhomogeneous reionization simulations: IR-A, IR-B, IR-C and IR-D (green, violet, orange and cyan lines respectively). We also show our fiducial flash reionization simulation for comparison (FR, dashed green lines). The top two panels show the evolution of the $T_0$ and $\gamma$ parameters describing the temperature-density relationship of the IGM (from top to bottom). It can be seen that while all models share the same $T_0$ evolution, they have slightly different $\gamma$ values at a given redshift, with the largest value corresponding to the flash model. We have checked that this is due to a larger scatter in temperature for a fixed density in the inhomogeneous models. Models that finished reionization at slightly later times (IR-B, IR-C, see Figure 3) have larger scatter which translates into smaller values fitted for the temperature-density slope, $\gamma$ (see discussion above). The bottom panel of Figure 9 presents the evolution of the pressure smoothing scale, $\lambda_{\rm P}$, as defined by Kulkarni et al. (2015). We see here a similar trend to the one found for the temperature-density slope, $\gamma$. Models that finish reionization at earlier times show a slightly higher value of $\lambda_{\rm P}$. As discussed above, the pressure smoothing scale is set by the full thermal history and not by the instantaneous temperature (Hui & Haiman 2003; Kulkarni et al. 2015).

In order to examine the possibility of distinguishing between these scenarios using Ly$\alpha$ forest statistics, we also computed the 1D flux power spectrum for these simulations following the same methodology as described in Section 4. Figure 10 shows the 1D flux power spectrum at redshifts $z = 6.0$, $z = 5.4$, $z = 5.0$ and $z = 4.2$. The models differ at large scales (low-$k$ modes), with the differences decreasing at lower redshifts. The bottom panels show the percentage difference between each model and our fiducial inhomogeneous simulation (IR-A). The largest-$k$ mode below which these differences are relevant also decreases as we go to lower redshifts. This is the same behavior as we found when we compared flash and inhomogeneous models (see Figure 7). These differences are due to the temperature fluctuations resulting from inhomogeneous reionization. In fact, we find that the differences between the inhomogeneous models are directly related to the duration of H I reionization, $\Delta z_{\rm reion,H\,I}$. The models with a more extended reionization (e.g. IR-B, see Figure 3) have more power at large scales. This is because the difference in temperature between the regions that reionized early and the ones that reionized last is larger (D'Aloisio et al. 2015). We found differences of up to $\sim 50\%$ in the power spectrum at $k \sim 10^{-3}$ s km$^{-1}$ between flash models and the most extended inhomogeneous reionization model (IR-B). While we are limited by the box size of our simulations ($L_{\rm box} = 40$ Mpc h$^{-1}$) these results clearly indicate that using the 1D flux power spectrum at high redshift could provide valuable constraints on the extent of reionization.

Figure 10 also shows the recent measurements of the 1D flux power spectrum computed by D'Aloisio et al. (2018b) at $z = 5.4$ using the 21 unique quasar spectra presented in (McGreer et al. 2015) (top right), Boera et al. (2018) at $z = 5.0$ (bottom left) and by Palanque-Delabrouille et al. (2013) at $z = 4.2$ (bottom right). A precise fit to these data is beyond the scope of this paper, as different degeneracies should be considered at the same time (e.g mean flux, cosmology, etc). However we think that a comparison with our various models illustrates that, based on the current observational precision and the differences that we find between our models, we should be able to distinguish between inhomogeneous reionization scenarios, especially considering global fits over a range of redshifts. Note that, in contrast to the need for accurate measurements of the cut-off of the 1D flux power spectrum (high-$k$ modes, e.g. Walther et al. 2018b), high-resolution measurements are not critical, and moderate-resolution spectra could provide the more precise measurement of the relevant low $k$ modes. In this regard, it is also worth mentioning that the data presented here is constitute only a small subset of all the data currently available.

Very interestingly, at small scales (high-$k$ modes) all the models show a very good agreement and are within $< 5\%$ of each other. This indicates that the cut-off of the 1D flux power spectrum is mainly sensitive to the heat input during reionization, $\Delta T$, and the median reionization redshift, $z_{\rm reion,H\,I}^{\rm median,inhomo}$. This is consistent with our findings comparing flash and inhomogeneous reionization models with different heat injection (see Figure 7). This similarity suggests that the heat input by H I reionization may be constrained from the Ly$\alpha$ forest with relatively simple flash models. Finally, we have not found any relevant features that indicate that the 1D flux power spectrum at $4 < z < 6$ is particularly sensitive to the different specific morphologies of the reionization models with different source populations, but this may change with larger simulation volumes.

# 6 TEMPERATURE AND DENSITY FIELDS AT HIGH-$Z$ IN REIONIZATION SIMULATIONS

In order to further investigate the relationship between the large-scale properties of the Ly$\alpha$ forest and the thermal properties of the IGM we studied in more detail the properties of the density and temperature fields of our simulations and the relationship between them.

Our first objective was to exclude the possibility that the large-scale power observed in the inhomogeneous runs is due to the scatter of the temperature-density relationship and is in fact due to large-scale (i.e. coherent) temperature fluctuations. For this purpose we started with the temperature field of the IR-A inhomogeneous reionization run at $z = 6$. We then generated a new temperature field by randomly shuffling the temperature values[7] in narrow density bins (width in log $\Delta$ of 0.05). As a result, this new painted temperature field that we obtained, shares the global thermal properties of the original one but has lost any spatial correlations with the large-scale density field. The two panels of Figure 11 illustrate this with a random slice of the original temperature field (left panel) and the same slice for the shuffled temperature field (right panel) at $z = 6$. The temperature-density relationship between these two models is indistinguishable by construction and therefore the $T_0$ and $\gamma$ parameters are identical. Because the two models also share the exact same density field, the gas pressure effects are also identical[8].

---

[7] We exclude cells with temperatures higher than log $T = 4.5$ K and densities higher than log $\Delta = 1.0$ from this analysis because we want to focus on the IGM properties and exclude shock-heated cells. In any case, the results presented here do not change if we include these cells.

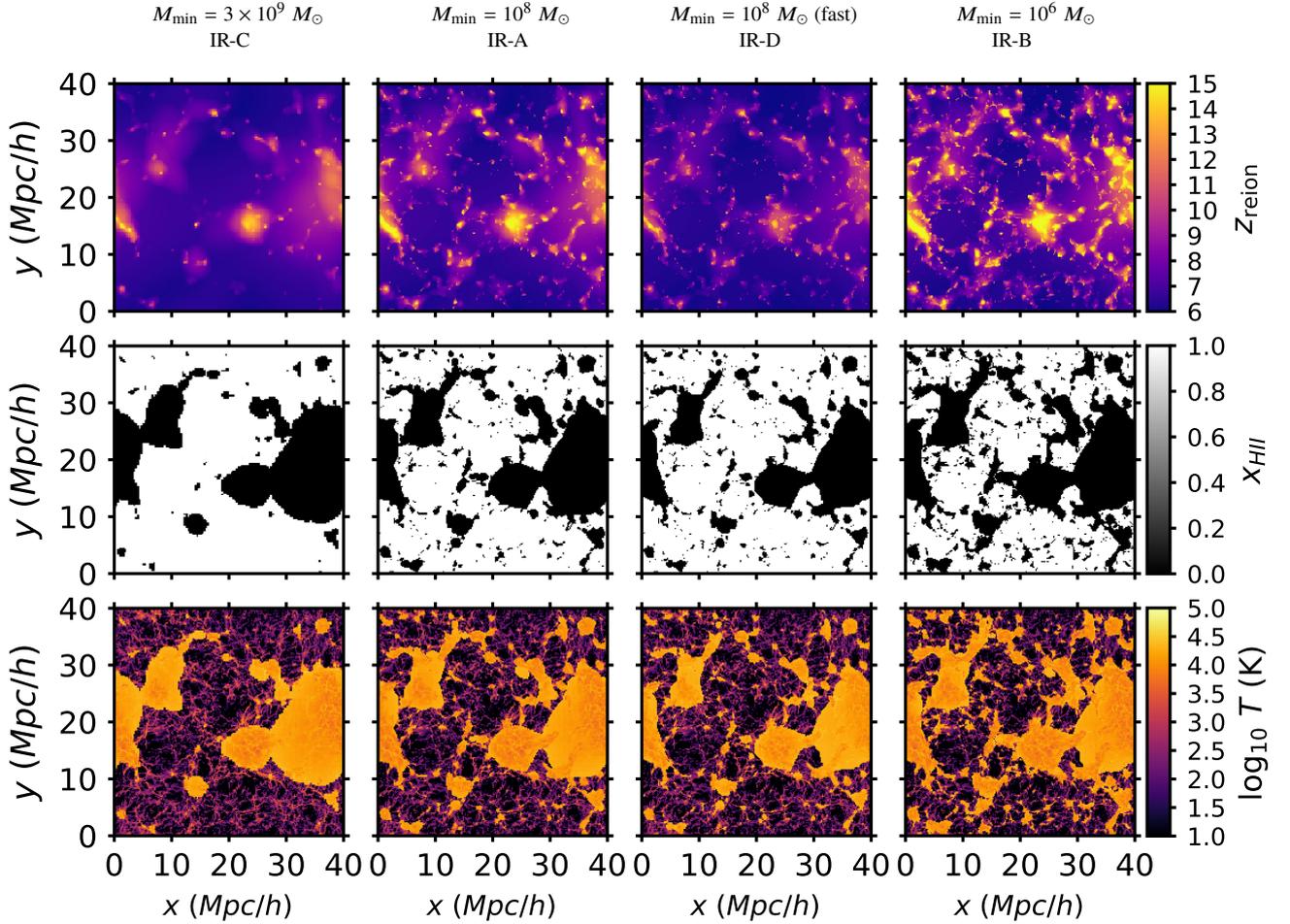

**Figure 8.** Top row: slices of the reionization redshift field for some of the inhomogeneous reionization simulations studied in this work, ordered from left to right by the minimum halo mass responsible for reionization: IR-C, IR-A, IR-D and IR-B. In these models we varied the dominant halo masses responsible for H I reionization ($3 \times 10^9 \, M_\odot$, $M_{\min} = 10^8$, $M_{\min} = 10^8$ (fast) and $10^6 \, M_\odot$ respectively) and adjusted the ionization efficiency so that they have a similar global reionization history (see Figure 3). As expected, reionization models dominated by smaller halo masses show morphologies with smaller ionized bubbles. Middle row: the same slices as in the top row, now showing the hydrogen ionized fraction for these simulations at $z = 8$ when they all have very similar overall ionization fraction. Bottom row: the same slices as above showing the temperature field at $z = 8$. It can clearly be seen how the typical ionized bubble size of the models directly translates into a different temperature scale.

We then computed the 1D flux power spectrum for the new model with the shuffled temperature field. We generated Ly$\alpha$ forest skewers using the new temperature field and the original density and velocity fields of the inhomogeneous run as well as the same UVB. We then renormalized the flux field to the observed mean flux at the respective redshift, as we did for the original simulation (see §4.1). The left panel of Figure 12 shows the 1D flux power spectrum at $z = 6$ of this new model (blue solid line) compared with the original inhomogeneous run (solid green line) and the flash run

(dashed green line). The differences at low-$k$ modes between the flash and inhomogeneous runs were discussed above in Section 4 (see also Figure 7), where we argued that they are due to the temperature fluctuations in the inhomogeneous reionization models. The 1D flux power spectrum of the model with the painted temperature field provides very interesting insights into the physical origin of the most relevant features of the 1D flux power spectrum, at both large and small scales. First, on large scales (low-$k$ modes) its shape seems to be more in agreement with the flash reionization model. The lack of large-scale power corroborates the idea that the extra power found at large scales is due to coherent temperature fluctuations. On small scales, the model with the shuffled temperatures shows significant extra power compared with both the flash and inhomogeneous models.

Because the overall normalization of the flux field obscures the effect of the different temperature fields, we also computed the 3D power spectrum of the temperature field fluctuations (i.e., $\delta T = T/\bar{T} - 1$) at $z = 6$ for these three simulations: the fiducial flash (FR) and inhomogeneous (IR-A) reionization simulations and the new model with the shuffled temperature. The right panel of Figure 12 shows these results. The first thing to notice is that the power spectrum obtained from the model in which we painted the temperature field has a very different overall power from the other two models. The power signal is reduced at all relevant scales and only increases at very large values ($k > 0.3$ s km$^{-1}$). Regarding the power spectrum at large scales ($k \lesssim 0.02$ s km$^{-1}$), only the fiducial inhomogeneous run, IR-A, presents a significant signal. This additional power comes from the temperature fluctuations resulting from inhomogeneous reionization and it is missed in both the flash model and the model in which we shuffled the temperature field. It is also interesting to focus on the small-scale differences between the models. The flash and inhomogeneous models agree quite well for $0.02 \lesssim k \lesssim 0.1$ s km$^{-1}$, confirming our conclusions from the analysis of the 1D flux power spectrum of these two type of models: for a similar heat injection during reionization the small-scale thermal properties are very similar as long as $z_{\rm reion, H\,I}^{\rm flash} = z_{\rm reion, H\,I}^{\rm median, inhomo}$. On top of the overall normalization mentioned above, the shape of the 3D temperature power spectrum at small scales ($0.02 \lesssim k \lesssim 0.1$ s km$^{-1}$) for the shuffled-temperature model seems to be steeper as we go to lower $k$ than those for the inhomogeneous or flash runs. Because we expect structure at these scales to be naturally suppressed, it is not surprising that by shuffling the temperature values at a fixed density from the inhomogeneous run we will introduce some extra power at small scales.

---

[8] Note, however, that $\lambda_P$ is not identical between runs, because in its definition the $n_{H\,I}$ field is used and this will not be the same in both runs. We will discuss this further below.

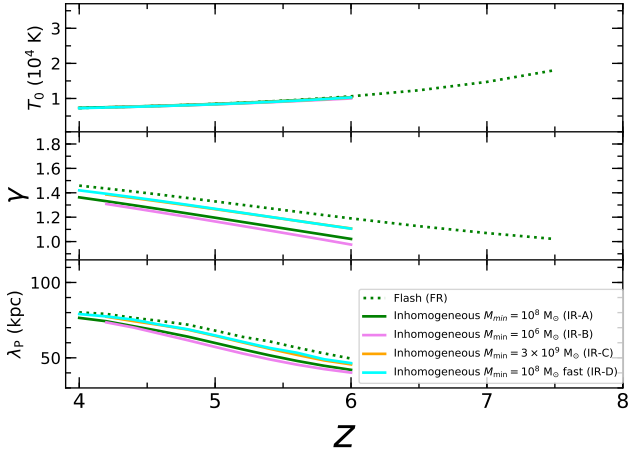

**Figure 9.** Thermal history for the inhomogeneous reionization simulations presented in this work: IR-A, IR-B, IR-C and IR-D. From top to bottom the panels show the temperature at mean density, $T_0$; the slope of the temperature-density relationship, $\gamma$; and the pressure smoothing scale, $\lambda_P$, as defined in Kulkarni et al. (2015).

We now want to further study how the density field is affected by the reionization process. Following the analysis that we carried out with the temperature field, we want to compute the 3D power spectrum of the gas density fluctuations field (i.e., $\delta\rho = \rho/\bar\rho - 1$). The gas power spectrum at small scales in cosmological simulations is dominated by high-density regions (see for example Kulkarni et al. 2015), however, and because we are interested in the properties of the IGM we computed the power spectrum clipping the density at different thresholds. This allowed us to explore the properties of the gas at lower densities corresponding to the IGM. We set all densities above the threshold to this exact value and then computed the 3D power spectrum of the gas density fluctuations of the resulting field (therefore $\bar\rho$ is different for each threshold). The left panel of

Figure 13 shows the 3D power spectrum of the gas density field fluctuations for the flash (FR) and inhomogeneous (IR-A) simulations at $z = 5$. We find that the gas density fields of the two simulations have almost identical properties. This confirms that the gas pressure effects in the two simulations are on average very similar, despite the different reionization histories.

The resulting 3D clipped gas power spectrum shows a clear cut-off which, we argue, describes the scale at which the pressure smoothing is relevant for that density. As we decrease the value of $\Delta_{\rm clip}$, this cut-off moves to lower-$k$ modes (i.e., the gas pressure scale increases) until we reach $\Delta_{\rm clip} \sim 1$, when the pressure scale seems to decrease again. This cut-off can be fitted by the function suggested in Kulkarni et al. (2015) for the pseudo real-space Ly$\alpha$ flux field:

$$\frac{k^3 P_{\rm gas}(k)^3}{2\pi^2} = A k^n \exp\left(-\frac{k^2}{k_\Delta^2}\right) \qquad (2)$$

which has three parameters: $A$, $n$ and $k_\Delta$. We define the scale associated with this cut-off as $\lambda_\Delta = 1/k_\Delta$. This naturally brings out the pressure smoothing scale - density relationship present in the density field (e.g. Gnedin & Hui 1998b). The right panel of Figure 13 shows the relationship between this cut-off scale and the clipping overdensity measured at $z = 5$ for the flash (FR, FRcold, FRhot, dashed lines) and inhomogeneous reionization (IR-A, IR-Acold, IR-Ahot, solid lines) simulations that we discussed in Section 4. As expected, models with a higher heat injection during reionization have a larger pressure scale, especially at densities below $\Delta \sim 10$. Interestingly, the analogous flash and inhomogeneous simulations, namely those share the same heat injection during reionization, have similar values at mean density. We think that the power spectrum of the smoothed gas density field could be a complementary and more detailed description of the gas pressure support in hydrodynamical simulations because it directly connects with the gas density field. It has the advantage that it is well defined at any redshift and it does not have any second-order dependence with the UVB or the temperature fields. A full detailed study of this possibility is beyond the scope of this paper but deserves theoretical consideration for future work as it could help to characterize in a much more accurate way the thermal properties of the IGM in hydrodynamical simulations.

## 7 UVB FLUCTUATIONS

Until now in our models we have considered a uniform UVB that evolves with redshift[9]. As discussed above, recent observations of the H I optical depth at redshift $z \gtrsim 5$ indicate that the H I photoionization and photoheating rates cannot be well described by uniform-field (see e.g. Becker et al. 2018). In this section we explore the effect that UVB fluctuations can have in the context of our new simulations.

We modelled fluctuations of the ionizing background in post-processing following the approach of Davies & Furlanetto (2016). In this approach we consider spatial variations of the mean free path, $\lambda_{\rm mfp}^{912}$, from modulations in the ionization state of optically thick absorbers assuming that $\lambda_{\rm mfp}^{912} \propto \Gamma_{\rm H\,I}^{2/3}/\Delta$, where $\Delta$ is the local matter density. We refer the reader to this paper for technical details.

Using this approach we created one model with a mean free path of $\langle \lambda_{\rm mfp}^{912} \rangle(z = 6) = 10$ cMpc. This is roughly a factor of 4 lower than the extrapolation of the measurements in Worseck et al. (2014) to this redshift and in this sense it can be considered an extreme model[10]. Our goal in this work, however, is to explore the possible effect that UVB fluctuations can have on the 1D flux power spectrum in the context of our new inhomogeneous and flash reionization models. We computed the fluctuating H I photoionization rate, $\Gamma_{\rm H\,I}$, at $z = 6$ on a uniform grid of $N = 64^3$, which resolves the large-scale fluctuations in the radiation field, and scaled this field up to $N = 2048^3$ (i.e., the same size as the hydrodynamical simulation) via linear interpolation in log space. The left panel of Figure 14 shows a slice of this field and shows that for this model the difference in the photoionization rates between regions can be as large as an order of magnitude. From this fluctuating UVB we computed a new neutral hydrogen density, $n_{\rm HI}$, for our fiducial flash (FR) and inhomogeneous reionization (IR-A) simulations using their original temperature and density fields for both simulations.

---

[9] In the inhomogeneous reionization simulations, if a resolution element is still not reionized the UVB is considered to be zero, but all resolution elements see the same UVB once they are reionized.

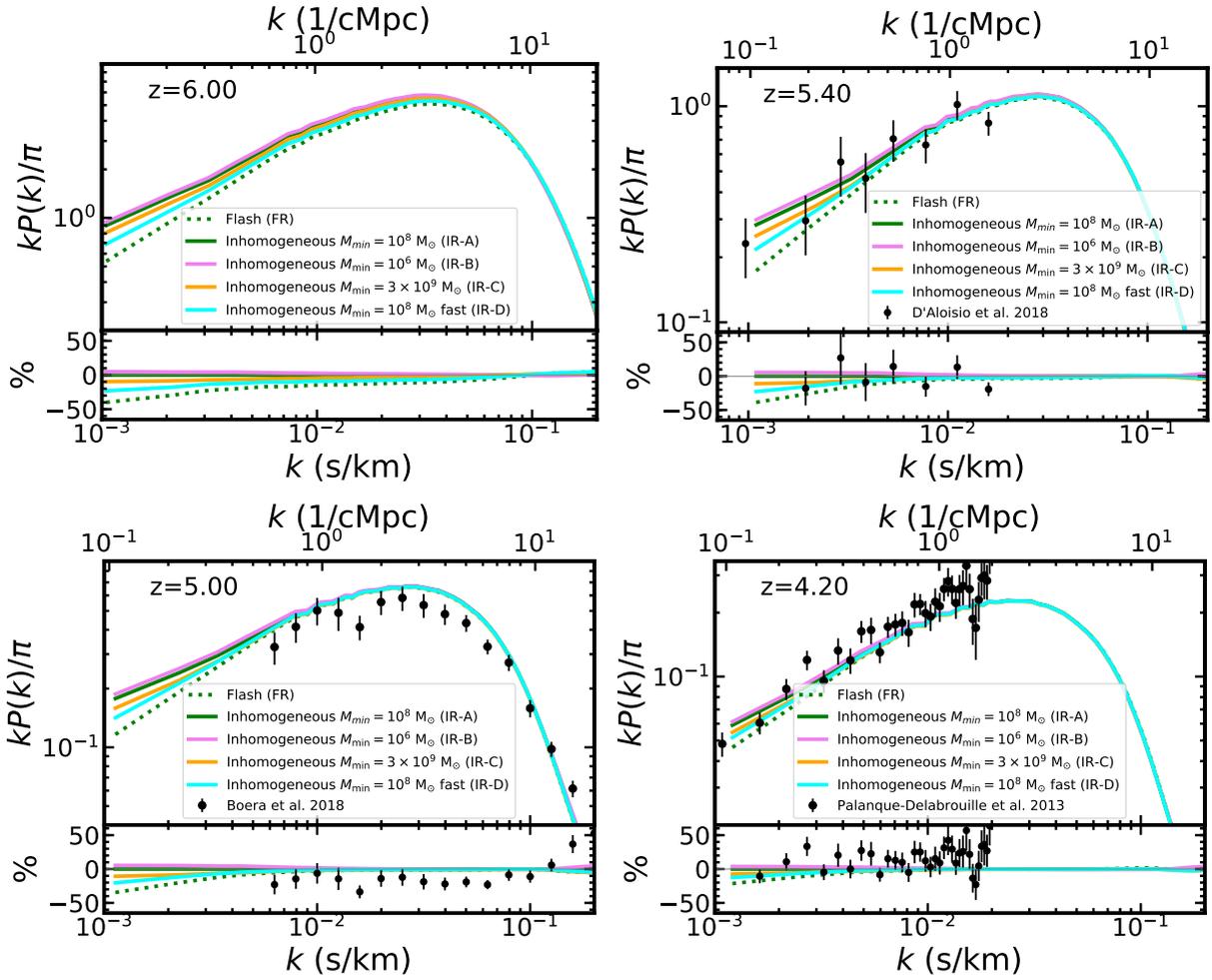

**Figure 10.** The 1D flux power spectrum at redshifts $z = 6.0$, $z = 5.4$, $z = 5.0$ and $z = 4.2$ for a set of different inhomogeneous reionization simulations: IR-A (green solid line), IR-B (magenta solid line), IR-C (orange solid line) and IR-D (cyan solid line) with the same heat input during reionization, $\Delta T = 2 \times 10^4$ K. In these models we have varied the dominant halo masses responsible for H I reionization ($M_{\rm min} = 10^8$, $10^6$ and $3 \times 10^9$ $M_\odot$ respectively) and adjusted the ionization efficiency so that they have a similar global reionization history (see Figure 3). We also show the 1D flux power spectrum for our FR flash reionization simulation, $z_{\rm reion, H\,I}^{\rm flash} = 7.75$, which also has the same heat input during reionization as the inhomogeneous runs. The bottom panels at each redshift show the percentage difference between each model and our fiducial inhomogeneous simulation (IR-A). Black circles and errorbars show observational measurements (see text for details).

Using the new neutral hydrogen density we computed the optical depth and the 1D flux power spectrum for each of these simulations. The right panel of Figure 14 compares the 1D flux power spectrum of these two new models with the original non-fluctuating ones. All models were normalized to the same mean flux. For the flash simulation, the effect seems to be an overall rescaling of the power spectrum at all scales. There is a small bump in power at low-$k$ ($k < 2 \times 10^{-3}$ s km$^{-1}$), however this scale is very close to the maximum scale measured by the simulation. The increase suggests that the effects of the UVB fluctuations could be more significant at even lower $k$ (larger scales), but we will need a larger simulation volume to study this effect further (D'Aloisio et al. 2018b). In the case of the modified inhomogeneous reionization simulation, however, the effects are clearer, with a change of the overall shape of the power spectrum. There is clear drop in the power at large scales (lower $k$) compared with the original model, indicating that the effect of thermal fluctuations in the power is cancelled by the photoionization fluctuations. This cancellation is a result of the opposite direction in which the two effects act to increase or decrease Ly$\alpha$ forest opacity on large scales (Davies & Furlanetto 2016; Davies et al. 2018a). UVB fluctuations tend to decrease the opacity in overdense regions with more sources of ionizing photons, and to increase the opacity in underdense regions that are distant from sources. Temperature fluctuations act in the opposite direction: overdense regions reionize early and are cold and opaque, while underdense regions reionize late and are hot and transparent. It is interesting to note that the UVB fluctuations do not seem to produce significant changes of the 1D flux power spectrum at small scales ($0.06 < k < 0.1$ s km$^{-1}$) for the inhomogeneous run. The effect in this range of scales is larger when the UVB fluctuations are applied to the flash reionization. The rise in power in the flash runs could be related to more dense/evolved parts of the Universe dominating the transmission. However, the reason for the similar power between the inhomogeneous runs at intermediate and small scales is not clear. This could be linked to the dispersion of the temperature-density relationship, indicating that the small-scale structure of the Ly$\alpha$ forest in models with a very tight relationship is much more sensitive to UVB fluctuations than in models that have more scatter about their temperature-density relationships.

---

[10] The box size of our simulations is too small to fully simulate strong UVB fluctuations, but we can approximate them by shrinking the mean free path.

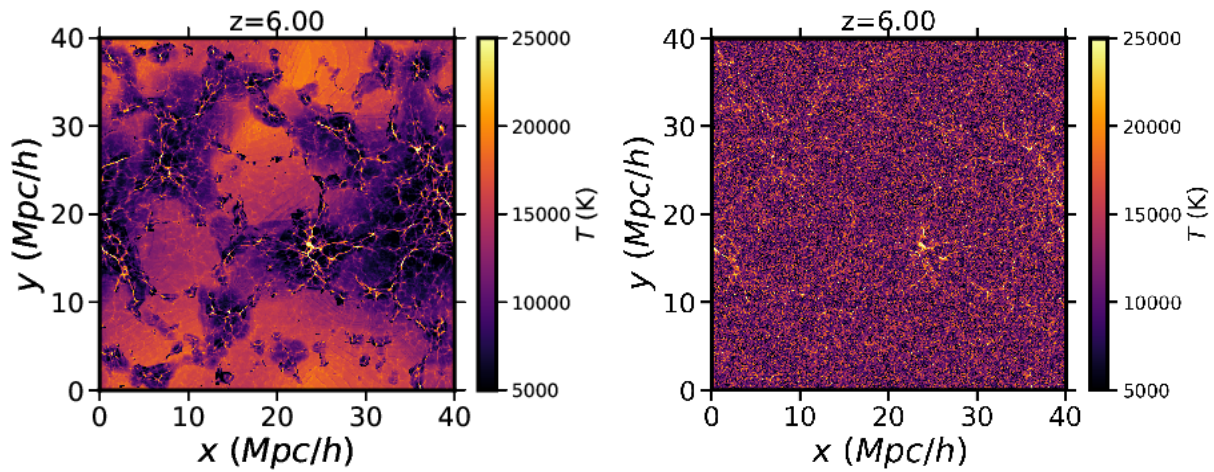

**Figure 11.** Painting the temperature field. The left panel shows a temperature slice of the inhomogeneous run (IR-A) at $z = 6$. The right panel shows a slice of the new painted temperature field constructed to reproduce the same temperature-density relationship as in the inhomogeneous run by shuffling its temperature values at fixed density. Although the two temperature-density relationships are indistinguishable by construction, the spatial correlation between the two fields is very different for each case. See text for more details.

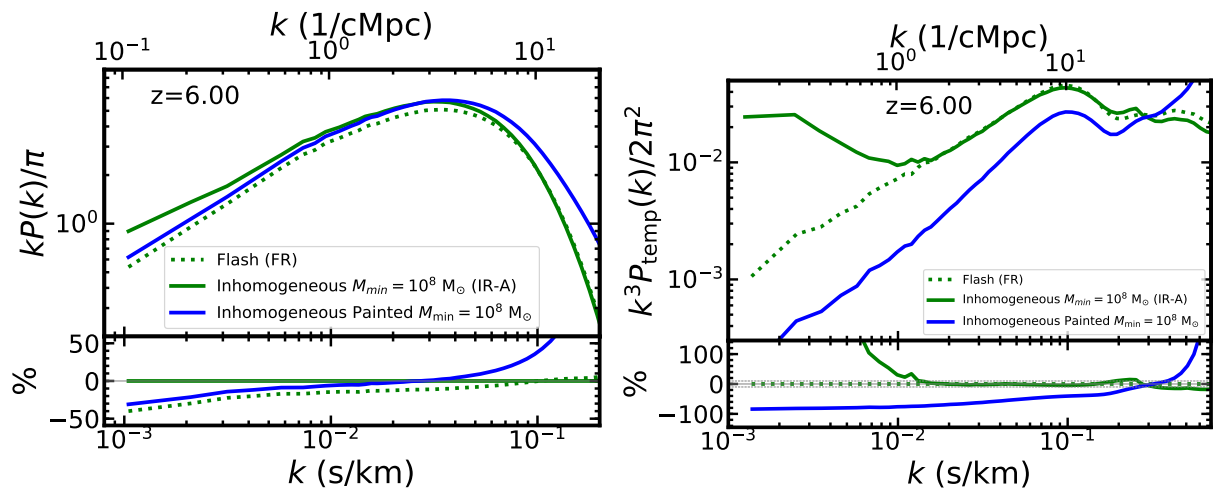

**Figure 12.** Temperature fields in reionization simulations. Left panel: the 1D flux power spectrum of our fiducial flash reionization simulation (FR, dashed green line), the inhomogeneous reionization simulation (IR-A, solid green line) and the painted temperature field (solid blue line). Right panel: The 3D power spectrum of the temperature field fluctuations (i.e., $\delta T = T/\bar{T} - 1$) for the same simulations.

## 8 DISCUSSION

In this work we have combined excursion set seminumerical models with cosmological hydrodynamical simulations in order to study the properties of the IGM and the Ly$\alpha$ forest to learn about H I reionization. The combination of box size ($L_{\rm box} = 40$ Mpc h$^{-1}$) and spatial resolution ($\Delta x \simeq 20$ kpc h$^{-1}$) chosen in our test runs has been driven mainly by the convergence constraints of the small scale Ly$\alpha$ forest properties. We will first discuss how this can limit

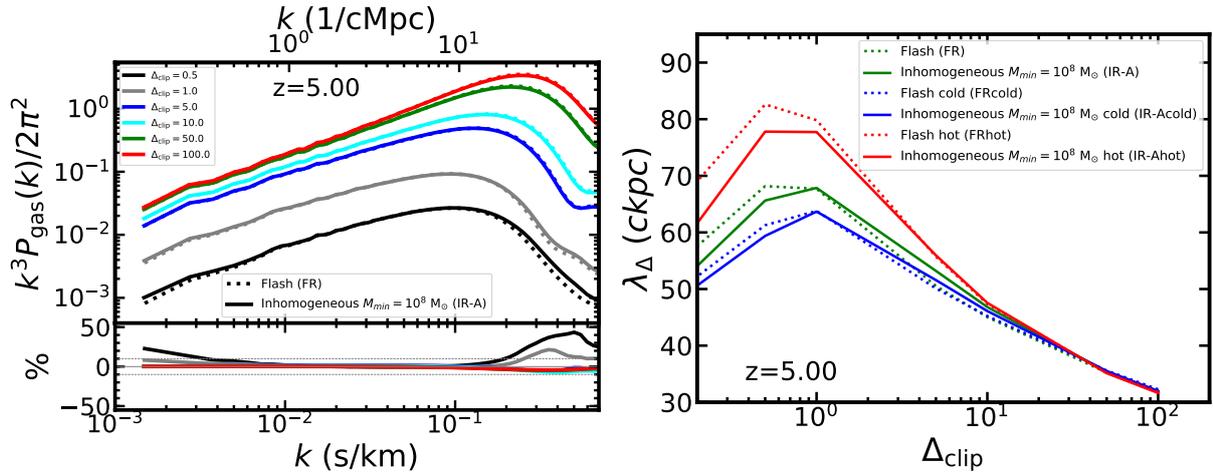

**Figure 13.** Left: the 3D power spectrum of the gas density field fluctuations (i.e., $\Delta\rho = \rho/\bar{\rho} - 1$) at $z = 5$ clipped at different density values for the FR flash reionization simulation (dashed lines) and the IR-A inhomogeneous reionization simulation (solid lines). The bottom panel shows the difference between the flash and inhomogeneous runs for each clipped density value. Right: the cut-off scale of the different clipped density fields, $\lambda_\Delta$, versus the density value used to clip the field. Our tests show that this relationship could help to quantify the gas pressure scale - density relationship in hydrodynamical simulations. See text for more details.

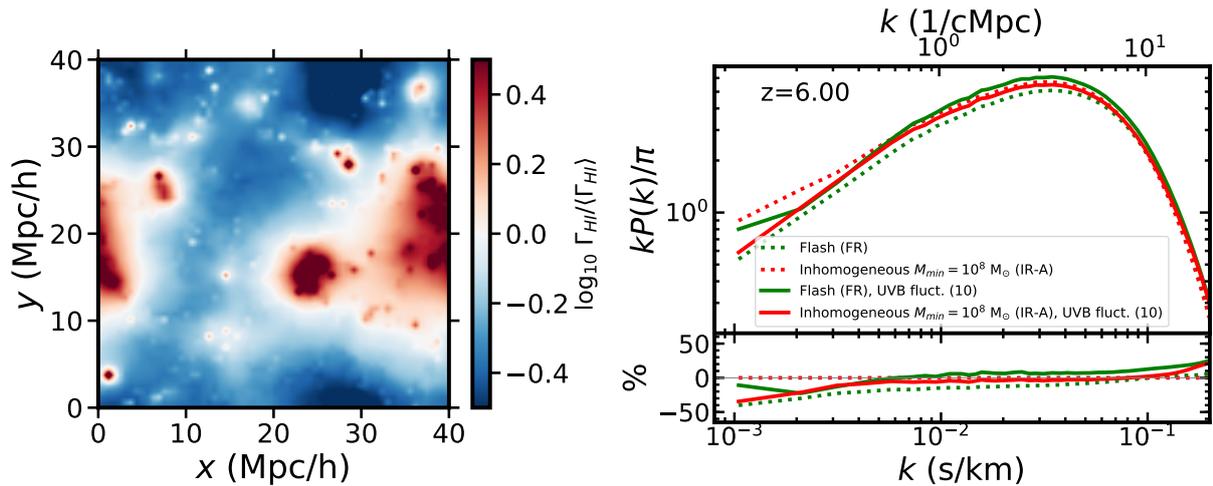

**Figure 14.** The addition of UV background (UVB) fluctuations in post-processing to our fiducial flash (FR) and inhomogeneous (IR-A) reionization models. Left panel: a slice showing the $\lambda_{\rm mfp} = 10$ Mpc model of the UVB fluctuation field used to post-process the hydrodynamical simulations. Right panel: the 1D flux power spectrum at $z = 6$ for the FR flash reionization model with and without UVB fluctuations (dashed and solid green lines, respectively), and the the IR-A inhomogeneous reionization model also with and without UVB fluctuations (dashed and solid red lines, respectively). See text for more details.

the methodology presented in this work as well as how it can affect the results.

The first relevant effect to be considered is the convergence of the reionization histories and field properties produced by our seminumerical approach that we used as input in our hydrodynamical simulations. Results from 3D radiative transfer cosmological simulations argued that box sizes of $L_{\rm box} \sim 100$ Mpc h$^{-1}$ or larger are needed to yield convergent H I reionization histories (e.g. Iliev et al. 2014). This is of course going to be related to the typical size of ionization bubbles in the reionization model, which is related to the sources responsible for reionization. The box size constraints cited above assume that galaxies in haloes of $M_h \sim 10^7 - 10^8\,M_\odot$ are the main drivers of reionization. From a purely computational perspective the goal of reaching $L_{\rm box} \sim 100$ Mpc h$^{-1}$ simulations while maintaining the necessary spatial resolution to resolve the Ly$\alpha$ forest is already doable. However, if larger halos were responsible for reionization, as in quasar-dominated scenarios, the box size needed for convergence will be larger. This is in fact currently the main limitation to using the method introduced in this work to simulate He II inhomogeneous reionization models in cosmological hydrodynamical simulations. He II reionization is driven by rare, luminous quasars (see e.g. Madau & Meiksin 1994) and therefore a much larger box is needed to obtain converged reionization models, $L_{\rm box} \gtrsim 500$ Mpc h$^{-1}$ (see e.g. Dixon et al. 2014; Davies et al. 2017; D'Aloisio et al. 2017). A short-term solution for this problem could be to use large-scale zoom-in cosmological simulations that can generate accurate H I and He II inhomogeneous reionization models and also Ly$\alpha$ forest predictions in a smaller subvolume.

In order to explore the effect of box size on the results from the hydrodynamical simulations, we ran the same flash and inhomoge-neous reionization models as presented in this work in a smaller box ($L_{box}$ = 20 Mpc h$^{-1}$) but using the same spatial resolution as in the larger boxes ($\Delta x \simeq 20$ kpc h$^{-1}$). We did not find any major difference from the main results shown in this work. We found convergence at the < 10% level for the 1D flux power spectrum at $4 < z < 6$[11] which agrees with similar recent tests using the Nyx code (Oñorbe et al. 2017b; Walther et al. 2018a). We refer to Lukić et al. (2015) for a more detailed discussion on resolution and box effects in the Ly$\alpha$ forest properties using Nyx. Similar results at these redshifts but for simulations using the Gadget code can be found in Bolton & Becker (2009) and Bolton et al. (2017).

There have been several studies to test the accuracy of seminumerical reionization models compared with radiative transfer simulations. These comparisons have shown that, despite the initial general idea that seminumerical methods will not be accurate for recreating the ionization history, because they do not chronologically follow the state of ionization at individual grid cells, they produce ionization histories that agree at the < 5% percentage level when compared with radiative transfer codes (Zahn et al. 2011; Majumdar et al. 2014; Hutter 2018). The agreement of the morphology of reionization between the two methods is however more complicated. Zahn et al. (2011) compared 100 Mpc h$^{-1}$ box-size 3D radiative transfer cosmological simulations with excursion set simulations and found a good agreement with the cross-correlation coefficient of the ionization fields in the range $0.6 < k < 1$ h Mpc$^{-1}$. The averaging of the density and ionizing emissivity fields in seminumerical methods results in more connected ionized regions, and strengthens the inside-out character of the ionization topology relative to radiative transfer simulations. More recent comparisons (Majumdar et al. 2014; Hutter 2018) between radiative transfer codes and new updated seminumerical methods, which used pre-run halo catalogues, have confirmed these results. By using higher resolution runs, however, Hutter (2018) found that the correlation between the two methods decreases towards smaller scales $k > 1$ h Mpc$^{-1}$ mainly for two main reasons. First, the more fractal shape of the ionization boundaries in the radiative transfer simulations leads to more small-scale structures in the ionization fields, and second, as reionization proceeds the ionized regions in the two methods merge at different points in space and time, and consequently develop increasingly different shapes with time. It is therefore necessary to perform a detailed comparison of the Ly$\alpha$ forest properties between 3D cosmological radiative transfer simulations with high enough spatial resolution and the method introduced in this work by using the reionization redshift field produced by the radiative transfer runs. We consider this to be beyond the scope of this paper, but we plan to undertake such a comparison in future work.

It is also interesting to consider the time-scale of the reionization process in the context of our method, because the spatial and time-step resolution used both in generating the reionization redshift fields and in the hydrodynamical simulations will set an upper limit on how well the exact morphology of the reionization models can be tracked in the simulation. Therefore we will now discuss the expected velocity of ionization fronts from previous works and then put that in the context of our results. First analytical estimates of the typical velocity of ionization fronts in the IGM during reionization gave values of $\gtrsim \sim 7$ h$^{-1}$ ckpc Myr$^{-1}$ ($\sim 10^4$ km s$^{-1}$, see e.g. Strömgren 1939). This is in good agreement with cosmological simulations with 3D radiative transfer that obtain velocities from $\sim 15$ h$^{-1}$ ckpc Myr$^{-1}$ ($\sim 2 \times 10^4$ km s$^{-1}$) around individual sources at first but accelerate up to $\sim 50$ h$^{-1}$ ckpc Myr$^{-1}$ ($\sim 7 \times 10^4$ km s$^{-1}$) as larger bubbles form (see e.g. Iliev et al. 2006; McQuinn et al. 2007; Trac & Cen 2007; Hirata 2018). Recently, Kaurov (2016) presented a detailed analysis of the time-scale of the reionization process using the results of a 3D radiative transfer cosmological simulation. These authors computed the time delay between the moments when each of the cells of the simulation (with a size of $\sim 156$ ckpc h$^{-1}$) reaches 10% and 90% ionization. They found a distribution of values that peaked at $\sim 20$ Myr (i.e., implying an ionization front velocity of $\sim 8$ h$^{-1}$ ckpc Myr$^{-1}$, $\sim 1 \times 10^4$ km s$^{-1}$), with most of the cells ranging between $\sim 10$ and $\sim 50$ Myr but with an extended tail to values as high as $\sim 500$ Myr[12]. While to generate the reionization redshift fields from the excursion set we used a very small time-step integration, $\Delta z = 0.05$, the spatial resolution, $\sim 230$ ckpc, is the current limiting factor (see Section 2.2). Regarding the hydrodynamical simulations, the spatial resolution of all runs presented in this work is $\sim 20$ ckpc/h and the typical time-step of these simulations during reionization is $\sim 1$ Myr[13]. Therefore the lower limit for the ionization-front velocity in our simulations is $\sim 20$ h$^{-1}$ ckpc Myr$^{-1}$ ($\sim 3 \times 10^4$ km s$^{-1}$). This is only a factor of $\sim 2$ larger than the fastest ionization-front velocities expected and therefore should be enough to capture the thermal structure of H I reionization.

Recently, D'Aloisio et al. (2018a) studied in detail the peak gas temperatures behind the ionization fronts, $T_{post-reion}$, using 1D high-resolution radiative transfer simulations. They found that $T_{post-reion}$ is only mildly sensitive to the spectrum of incident radiation over most of the parameter space, with temperatures set primarily by the ionization-front speeds. They measured the velocity of ionization fronts in 3D cosmological radiative transfer SCORCH simulations (Doussot et al. 2017), finding also a broad range of velocities during the early phases of reionization, from $\sim 0.05$ to $\sim 1.5$ h$^{-1}$ ckpc Myr$^{-1}$ (50 to $2 \times 10^3$ km s$^{-1}$). However, the velocity increases to $\sim 7$ h$^{-1}$ ckpc Myr$^{-1}$ ($\sim 10^4$ km s$^{-1}$), and the distribution narrows by the end of H I reionization. Mapping these velocities to temperatures yields $T^{post-reion} = 1.7 \times 10^3 - 2.2 \times 10^4$ K during the first half of reionization, but hotter temperatures of $T^{post-reion} = 2.5 \times 10^3 - 3.0 \times 10^4$ K can be reached during the very last phases of reionization. These findings are extremely interesting in the context of this work and they indicate that assuming just one value of $T^{post-reion}$ during reionization, as we have done in our simulations, could be too simplistic. These findings are easy to implement in the context of the method introduced here, however, and therefore also allow for a more accurate, but cheap, modelling of the process in cosmological hydrodynamical simulations. We definitely plan to explore this issue in detail in the near future.

Finally, in order to further study the effects on small-scale Ly$\alpha$ statistics between our different reionization models, we also calculated the curvature flux statistics (Becker et al. 2011). We have confirmed that all results presented in this work for the shape of the 1D flux power spectrum cut-off regarding the comparison between different reionization models hold also for the curvature statistics.

---

[11] This is the convergence level at the highest $k$ mode studied, 0.1 s km$^{-1}$, but it diminishes as we go to lower $k$ values.

[12] The time intervals between the snapshots limited the lowest value that could be measured to $\sim 6$ Myr but this does not affect the main results.

[13] This is assuming Courant factors of $\sim 0.2 - 0.5$.

# 9 CONCLUSIONS

In this paper we have introduced a new method to model inhomogeneous reionization scenarios in optically thin cosmological hydrodynamical simulations that does not increase their cost. In this method, each resolution element in the simulation is assigned its own reionization redshift. To account for the heating produced by the ionization front during the first time-step of the simulation, in which the reionization redshift is crossed, a fixed amount of energy, $\Delta T$, is injected, which is a free parameter in the model. From the first time-step after crossing the reionization redshift and onwards, the resolution element is assumed to be optically thin and can see the homogeneous UVB.

Using this method we produced a suite of $2048^3$ and $L_{\rm box} = 40$ Mpc h$^{-1}$ Nyx cosmological hydrodynamical simulations with different reionization models parametrized by the midpoint of reionization, $z_{\rm reion,H\,I}^{\rm median}$, its duration, $\Delta z_{\rm reion,H\,I}$, and the heat injection, $\Delta T$. To generate different redshift reionization models we have used seminumerical methods that combine the excursion set formalism with the initial conditions of the simulation and approximate radiative transfer methods (e.g. 21cmFAST, Mesinger et al. 2011), which allowed us to efficiently explore the parameter space. In order to make a comparison with the inhomogeneous models, we also ran a set of instantaneous or flash reionization models in which we fixed all resolution elements in the simulation to the same reionization redshift.

We first analysed the temperature and density fields and their relationship for all these simulations during and after reionization. As expected, inhomogeneous reionization models enhance the temperature fluctuations found in the IGM once reionization is completed, as regions that reionized early have had time to cool while regions that reionize at late times are still warmer. For this reason reionization models with the same heat injection but that occurred over a longer period of time produce larger temperature fluctuations (D'Aloisio et al. 2018b). Temperature fluctuations alter the temperature-density relationship in the IGM after reionization, significantly increasing its scatter compared with flash reionization models.

To study the effects of these different reionization models on the properties of the Ly$\alpha$ forest, we computed the Ly$\alpha$ 1D flux power spectrum at $4 < z < 6$, namely after reionization is finished, for our simulation ensemble. We found that thermal fluctuations can produce a significant increase of power (up to $\sim 50\%$) at large scales, $k \lesssim 0.01$ s km$^{-1}$. Reionization models with larger thermal injection have more power at these scales with the differences decreasing with redshift as the fluctuations decrease. We quantified these differences for our suite of simulations and showed that available observational data could already provide interesting constraints on the duration of H I reionization.

We confirmed that the small-scale differences in the 1D power spectrum (high-$k$ modes) are driven by the thermal history of the IGM, and therefore by studying them we can constrain when H I reionization happened and how much heat was injected (Nasir et al. 2016; Oñorbe et al. 2017b; Boera et al. 2018). Moreover, we found that once reionization is finished, the small-scale properties of the Ly$\alpha$ forest in inhomogeneous and flash reionization are very similar provided the models share the same heat injection during reionization, $\Delta T$, and the reionization in the flash models occurs at the median reionization redshift of the extended model, that is, $z_{\rm reion,H\,I}^{\rm flash} = z_{\rm reion,H\,I}^{\rm median,inhomo}$. This result indicates that there is a scale below which the correlations resulting from inhomogeneous reionization average out and that the the Ly$\alpha$ forest properties are well described by an average of the density-dependent thermal properties on these scales. We have shown that this scale could be as large as $k \gtrsim 0.02$ s km$^{-1}$ at $4 < z < 6$ and therefore relevant for available observations and for constraining not only reionization but also other physical processes relevant at these scales (e.g. the nature of dark matter).

To further investigate these findings we compared the properties of the density and temperature fields of the different reionization models. The spatial correlations of these fields at large and small scales show a clear correlation with the Ly$\alpha$ forest properties. We confirmed that the difference in the large-scale power between the inhomogeneous and flash models is indeed due to these correlations and not to the larger scatter in the temperature-density relationship that naturally arises in the inhomogeneous models. Moreover, we showed that the spatial correlations of the density and temperature fields and their relationship are crucial to modelling the small scales of the Ly$\alpha$ forest properly. In fact, we found that the classical paradigm of describing the thermal state of the IGM with three parameters, two for the temperature-density relationship, plus one describing the pressure scale of the gas, does not fully capture the differences between models, once inhomogeneous reionization models are considered with high enough precision.

We also studied the effects of UVB fluctuations in our simulations to see how they could affect the Ly$\alpha$ forest. We constructed an extreme model ($\lambda_{\rm mfp} = 10$ Mpc) that we applied in post-processing to our fiducial flash and inhomogeneous reionization simulations and compared the 1D flux power spectrum between all these models. We found that UVB fluctuations alter the overall normalization of the power spectrum by changing the density-optical depth mapping, producing effects on all scales. It is interesting however that the UVB fluctuations seem to cancel the large-scale effects that appear in the inhomogeneous reionization model owing to thermal fluctuations. In any case these results indicate that this effect should be also considered when using the 1D flux power spectrum at high $z$ to constrain H I reionization, cosmological parameters and/or the nature of dark matter. It is worth noting that thermal fluctuations will last long past the end of reionization, while UVB fluctuations will decay faster; therefore, in principle, these effects could be disentangled.

Finally, we believe that the new method introduced in this work is an important step forward to accurately modelling the H I and He II reionization processes in cosmological hydrodynamical simulations. The method is easy to implement and adds no significant extra computing cost compared with the standard approach used in IGM and galaxy formation and evolution studies, which assumes a simple homogeneous reionization model. We plan to further explore the limitations of this approach both by implementing more accurate thermal heating models and by comparing it with high-resolution 3D radiative transfer cosmological simulations.

# ACKNOWLEDGEMENTS

We thank the members of the ENIGMA group at the Max Planck Institute for Astronomy (MPIA) for helpful discussions. Calculations presented in this paper used the COBRA and DRACO cluster of the Max Planck Computing and Data Facility (MPCDF, formerly known as RZG) MPCDF is a competence centre of the Max Planck Society located in Garching (Germany). We acknowledge PRACE for awarding us access to JUWELS hosted by JSC at Jülich (Germany). We also used resources of the National Energy Research Scientific Computing Center (NERSC), which is supported by the Office of

**REFERENCES**

Abel T., Haehnelt M. G., 1999, ApJ, 520, L13
Almgren A. S., Bell J. B., Lijewski M. J., Lukić Z., Van Andel E., 2013, ApJ, 765, 39
Alvarez M. A., Finlator K., Trenti M., 2012, ApJ, 759, L38
Bañados E., et al., 2014, AJ, 148, 14
Bañados E., et al., 2018, Nature, 553, 473
Battaglia N., Trac H., Cen R., Loeb A., 2013, ApJ, 776, 81
Bauer A., Springel V., Vogelsberger M., Genel S., Torrey P., Sijacki D., Nelson D., Hernquist L., 2015, MNRAS, 453, 3593
Becker G. D., Bolton J. S., 2013, MNRAS, 436, 1023
Becker G. D., Bolton J. S., Haehnelt M. G., Sargent W. L. W., 2011, MNRAS, 410, 1096
Becker G. D., Bolton J. S., Madau P., Pettini M., Ryan-Weber E. V., Venemans B. P., 2015, MNRAS, 447, 3402
Becker G. D., Davies F. B., Furlanetto S. R., Malkan M. A., Boera E., Douglass C., 2018, ApJ, 863, 92
Boera E., Becker G. D., Bolton J. S., Nasir F., 2018, preprint, (arXiv:1809.06980)
Bolton J. S., Becker G. D., 2009, MNRAS, 398, L26
Bolton J. S., Puchwein E., Sijacki D., Haehnelt M. G., Kim T.-S., Meiksin A., Regan J. A., Viel M., 2017, MNRAS, 464, 897
Bond J. R., Cole S., Efstathiou G., Kaiser N., 1991, ApJ, 379, 440
Bosman S. E. I., Fan X., Jiang L., Reed S., Matsuoka Y., Becker G., Haehnelt M., 2018, MNRAS, 479, 1055
Cen R., McDonald P., Trac H., Loeb A., 2009, ApJ, 706, L164
Chardin J., Haehnelt M. G., Aubert D., Puchwein E., 2015, MNRAS, 453, 2943
Chardin J., Puchwein E., Haehnelt M. G., 2017, MNRAS, 465, 3429
Choudhury T. R., Puchwein E., Haehnelt M. G., Bolton J. S., 2015, MNRAS, 452, 261
Coc A., Uzan J.-P., Vangioni E., 2013, preprint, (arXiv:1307.6955)
D'Aloisio A., McQuinn M., Trac H., 2015, ApJ, 813, L38
D'Aloisio A., Upton Sanderbeck P. R., McQuinn M., Trac H., Shapiro P. R., 2017, MNRAS, 468, 4691
D'Aloisio A., McQuinn M., Maupin O., Davies F. B., Trac H., Fuller S., Upton Sanderbeck P. R., 2018a, preprint, (arXiv:1807.09282)
D'Aloisio A., McQuinn M., Davies F. B., Furlanetto S. R., 2018b, MNRAS, 473, 560
Davies F. B., Furlanetto S. R., 2016, MNRAS, 460, 1328
Davies F. B., Furlanetto S. R., McQuinn M., 2016, MNRAS, 457, 3006
Davies F. B., Furlanetto S. R., Dixon K. L., 2017, MNRAS, 465, 2886
Davies F. B., Becker G. D., Furlanetto S. R., 2018a, ApJ, 860, 155
Davies F. B., et al., 2018b, ApJ, 864, 142
Dayal P., Ferrara A., 2018, Phys. Rep., 780, 1
Dixon K. L., Furlanetto S. R., Mesinger A., 2014, MNRAS, 440, 987
Doussot A., Trac H., Cen R., 2017, preprint, (arXiv:1712.04464)
Eilers A.-C., Davies F. B., Hennawi J. F., 2018, ApJ, 864, 53
Fan X., et al., 2006, AJ, 132, 117
Faucher-Giguère C.-A., Lidz A., Zaldarriaga M., Hernquist L., 2009, ApJ, 703, 1416
Feng Y., Di-Matteo T., Croft R. A., Bird S., Battaglia N., Wilkins S., 2016, MNRAS, 455, 2778
Finlator K., Keating L., Oppenheimer B. D., Davé R., Zackrisson E., 2018, MNRAS, 480, 2628
Furlanetto S. R., Oh S. P., 2009, ApJ, 701, 94
Furlanetto S. R., Zaldarriaga M., Hernquist L., 2004, ApJ, 613, 1
Giallongo E., et al., 2015, A&A, 578, A83
Gnedin N. Y., 2014, ApJ, 793, 29
Gnedin N. Y., Hui L., 1998a, MNRAS, 296, 44
Gnedin N. Y., Hui L., 1998b, MNRAS, 296, 44
Gnedin N. Y., Becker G. D., Fan X., 2017, ApJ, 841, 26
Greig B., Mesinger A., Haiman Z., Simcoe R. A., 2017, MNRAS, 466, 4239
Gunn J. E., Peterson B. A., 1965, ApJ, 142, 1633
Haardt F., Madau P., 1996, ApJ, 461, 20
Haardt F., Madau P., 2001, in Neumann D. M., Tran J. T. V., eds, Clusters of Galaxies and the High Redshift Universe Observed in X-rays. p. 64 (arXiv:astro-ph/0106018)
Haardt F., Madau P., 2012, ApJ, 746, 125
Hahn O., Abel T., 2011, MNRAS, 415, 2101
Hassan S., Davé R., Finlator K., Santos M. G., 2016, MNRAS, 457, 1550
Hirata C. M., 2018, MNRAS, 474, 2173
Hopkins P. F., et al., 2018, MNRAS, 480, 800
Howlett C., Lewis A., Hall A., Challinor A., 2012, Journal of Cosmology and Astroparticle Physics, 4, 27
Hui L., Haiman Z., 2003, ApJ, 596, 9
Hutter A., 2018, MNRAS, 477, 1549
Iliev I. T., Mellema G., Pen U.-L., Merz H., Shapiro P. R., Alvarez M. A., 2006, MNRAS, 369, 1625
Iliev I. T., et al., 2009, MNRAS, 400, 1283
Iliev I. T., Mellema G., Ahn K., Shapiro P. R., Mao Y., Pen U.-L., 2014, MNRAS, 439, 725
Iršič V., et al., 2017, Phys. Rev. D, 96, 023522
Katz N., Weinberg D. H., Hernquist L., 1996, ApJS, 105, 19
Kaurov A. A., 2016, ApJ, 831, 198
Keating L. C., Puchwein E., Haehnelt M. G., 2018, MNRAS, 477, 5501
Khaire V., Srianand R., 2018, preprint, (arXiv:1801.09693)
Kim H.-S., Wyithe J. S. B., Park J., Poole G. B., Lacey C. G., Baugh C. M., 2016, MNRAS, 455, 4498
Kulkarni G., Hennawi J. F., Oñorbe J., Rorai A., Springel V., 2015, ApJ, 812, 30
Kulkarni G., Choudhury T. R., Puchwein E., Haehnelt M. G., 2016, MNRAS, 463, 2583
Kulkarni G., Keating L. C., Haehnelt M. G., Bosman S. E. I., Puchwein E., Chardin J., Aubert D., 2018, preprint, (arXiv:1809.06374)
La Plante P., Trac H., Croft R., Cen R., 2017, ApJ, 841, 87
Lee K.-G., Suzuki N., Spergel D. N., 2012, AJ, 143, 51
Lewis A., Challinor A., Lasenby A., 2000, ApJ, 538, 473
Lidz A., Malloy M., 2014, ApJ, 788, 175
Lidz A., McQuinn M., Zaldarriaga M., Hernquist L., Dutta S., 2007, ApJ, 670, 39
Liu A., Pritchard J. R., Allison R., Parsons A. R., Seljak U., Sherwin B. D., 2016, Phys. Rev. D, 93, 043013
Lukić Z., Stark C. W., Nugent P., White M., Meiksin A. A., Almgren A., 2015, MNRAS, 446, 3697
Madau P., Meiksin A., 1994, ApJ, 433, L53
Majumdar S., Mellema G., Datta K. K., Jensen H., Choudhury T. R., Bharadwaj S., Friedrich M. M., 2014, MNRAS, 443, 2843
McGreer I. D., Mesinger A., D'Odorico V., 2015, MNRAS, 447, 499
McQuinn M., 2012, MNRAS, 426, 1349
McQuinn M., 2016, ARA&A, 54, 313
McQuinn M., Upton Sanderbeck P. R., 2016, MNRAS, 456, 47
McQuinn M., Lidz A., Zahn O., Dutta S., Hernquist L., Zaldarriaga M., 2007, MNRAS, 377, 1043
Meiksin A., Tittley E. R., 2012, MNRAS, 423, 7
Mesinger A., Furlanetto S., Cen R., 2011, MNRAS, 411, 955
Mesinger A., Aykutalp A., Vanzella E., Pentericci L., Ferrara A., Dijkstra M., 2015, MNRAS, 446, 566
Miralda-Escudé J., Rees M. J., 1994, MNRAS, 266, 343
Mortlock D. J., et al., 2011, Nature, 474, 616
Nasir F., Bolton J. S., Becker G. D., 2016, MNRAS, 463, 2335
Norman M. L., Reynolds D. R., So G. C., Harkness R. P., Wise J. H., 2015, ApJS, 216, 16
Oñorbe J., Hennawi J. F., Lukić Z., 2017a, ApJ, 837, 106
Oñorbe J., Hennawi J. F., Lukić Z., Walther M., 2017b, ApJ, 847, 63
Ocvirk P., et al., 2016, MNRAS, 463, 1462

Palanque-Delabrouille N., et al., 2013, A&A, 559, A85
Park H., Shapiro P. R., Choi J.-h., Yoshida N., Hirano S., Ahn K., 2016, ApJ, 831, 86
Pawlik A. H., Schaye J., Dalla Vecchia C., 2015, MNRAS, 451, 1586
Peeples M. S., Weinberg D. H., Davé R., Fardal M. A., Katz N., 2010, MNRAS, 404, 1281
Pillepich A., et al., 2018, MNRAS, 473, 4077
Planck Collaboration et al., 2018, preprint, (arXiv:1807.06205)
Poole G. B., Angel P. W., Mutch S. J., Power C., Duffy A. R., Geil P. M., Mesinger A., Wyithe S. B., 2016, MNRAS, 459, 3025
Press W. H., Schechter P., 1974, ApJ, 187, 425
Puchwein E., Haardt F., Haehnelt M. G., Madau P., 2018, preprint, (arXiv:1801.04931)
Robertson B. E., Ellis R. S., Furlanetto S. R., Dunlop J. S., 2015, ApJ, 802, L19
Schaye J., et al., 2015, MNRAS, 446, 521
So G. C., Norman M. L., Reynolds D. R., Wise J. H., 2014, ApJ, 789, 149
Sobacchi E., Mesinger A., 2014, MNRAS, 440, 1662
Strömgren B., 1939, ApJ, 89, 526
Tittley E. R., Meiksin A., 2007, MNRAS, 380, 1369
Trac H., Cen R., 2007, ApJ, 671, 1
Trac H., Cen R., Loeb A., 2008, ApJ, 689, L81
Venemans B. P., et al., 2015a, MNRAS, 453, 2259
Venemans B. P., et al., 2015b, ApJ, 801, L11
Viel M., Haehnelt M. G., Springel V., 2004, MNRAS, 354, 684
Viel M., Becker G. D., Bolton J. S., Haehnelt M. G., 2013a, Phys. Rev. D, 88, 043502
Viel M., Schaye J., Booth C. M., 2013b, MNRAS, 429, 1734
Walther M., Oñorbe J., Hennawi J. F., Lukić Z., 2018a, preprint, (arXiv:1808.04367)
Walther M., Hennawi J. F., Hiss H., Oñorbe J., Lee K.-G., Rorai A., O'Meara J., 2018b, ApJ, 852, 22
Wang F., et al., 2017, ApJ, 839, 27
Weigel A. K., Schawinski K., Treister E., Urry C. M., Koss M., Trakhtenbrot B., 2015, MNRAS, 448, 3167
Willott C. J., et al., 2010, AJ, 139, 906
Wise J. H., Demchenko V. G., Halicek M. T., Norman M. L., Turk M. J., Abel T., Smith B. D., 2014, MNRAS, 442, 2560
Worseck G., et al., 2014, MNRAS, 445, 1745
Zahn O., Mesinger A., McQuinn M., Trac H., Cen R., Hernquist L. E., 2011, MNRAS, 414, 727
Zaldarriaga M., Hui L., Tegmark M., 2001, ApJ, 557, 519

# APPENDIX A: INHOMOGENOUS REIONIZATION FIELDS AND HEAT INJECTION: TESTS

In this Appendix we want to address how different assumptions in the creation of the initial reionization field and how the heat due to reionization is injected can affect the results presented in this paper, as well how exactly the heat injection is implemented in the hydrodynamical simulation. All simulations discussed here have a box-size length of $L_{box} = 20$ Mpc h$^{-1}$ and $1024^3$ resolution elements. Therefore, although they cover a smaller volume than the fiducial runs, they do have the same spatial resolution.

We first want to modify how the heat due to reionization is injected in the simulation. In our default implementation we have considered that there is a maximum temperature that any resolution element can reach after reionization which corresponds to the free parameter, $\Delta T$. We ran a simulation (IR-A-noTthres) modifying how the final temperature of each resolution element, $T_{post-reion}$, is computed in the simulation by removing the condition of setting an upper limit of the final temperature equal to $\Delta T$. This basically just translates to changing Equation 1 to:

$$T_{post-reion} = x_{H\,I,pre-reion}\Delta T + T_{pre-reion} \quad (A1)$$

where $T_{pre-reion}$ is the temperature of the resolution element just before reionization and $x_{H\,I}$ is its neutral fraction also before reionization. Therefore resolution elements can now in fact be above the temperature injected due to reionization, and a slightly larger scatter in temperatures is produced. The exact parameters of this simulation can be found in Table 1. The top panel of Figure A1 shows a slice of the temperature field at $z = 6$ of a simulation run using our fiducial model (IR-A). The left panel in the second row shows the same slice without imposing any temperature threshold. Both slices are very similar; however, there are some cells that are reionized at very late times in the simulation that in the fiducial simulation have lower temperature values (e.g. in the lower left part of the plot). However, as the left panel of Figure A2 shows for $z = 6$, when the 1D flux power spectra for both models are compared they do not show any significant differences at scales reached by observations ($k \lesssim 0.02$ s km$^{-1}$). The right panel of Figure A2 shows the 3D power spectrum of the fluctuations of the temperature field ($\delta T = T/\bar{T} - 1$) for these runs. Note that, despite having slightly different mean temperature values, the two simulations show a very similar 3D temperature power spectrum at all scales.

In order to address the effects of the time resolution used when generating the reionization models we created a new model (IR-A-dz) in which we used a higher time resolution $\Delta z = 0.01$ than in our default runs ($\Delta z = 0.05$). The exact parameters of this simulation can be found in Table 1. We first confirmed that using a different time resolution did not change the reionization history of the model. Note that, as in the previous test, in this simulation we used Equation (A1) to calculate the temperature after reionization (i.e. we did not consider a temperature threshold). The right panel in the second row of Figure A1 shows a slice of the temperature field at $z = 6$ of the new model with higher resolution time. While the effect is very subtle the temperature map is smoother in the run with a higher time resolution. Figure A2 shows the 1D flux power spectrum at $z = 6$ for this model (left panel) and the 3D temperature power spectrum (right panel) and demonstrates that the differences between the two simulations are very small and only relevant above scales of $k > 0.2$ s km$^{-1}$, above the modes in which we are interested in this work. The differences in the temperature field are in agreement with the test run in which we only modified the reionization heating (see above).

Finally, the bottom row panels of Figure A1 show the effects of using different smoothing algorithms to generate the high-resolution reionization map used as input in the simulations from the one obtained using our seminumerical approach. Our fiducial model (top panel) does not post-process this field, but we explored the effects of applying a simple linear interpolation scheme (IR-A-li, bottom row left panel) and a gaussian smoothing approach (IR-A-gs, bottom row right panel). Notice that in these two simulations we also used equation (A1) to calculate the temperature after reionization (i.e. we did not consider a temperature threshold). The temperature slices show very clearly the effect of smoothing the reionization field generating a smoother temperature field by $z = 6$.

Figure A2 shows the 1D flux power spectrum (left panel) and the 3D temperature power spectrum (right panel) at $z = 6$ for the two smoothed models (dotted orange and red lines for the linear and gaussian interpolation respectively). While the flux power spectrum shows that the differences from our fiducial run are not relevant at the scales studied in this work ($k < 0.1$ s km$^{-1}$) they have a substantial effect at smaller scales. This is even clearer when we look at the 3D temperature power spectrum, as at small scales the smoothing clearly changes its overall shape. Therefore we decided not to smooth any reionization field in our fiducial runs. Although 22  *J. Oñorbe et al.***Table 1.** Summary of simulations used to test the reionization field.

| Sim | Method | $z_{\text{reion,H\,I}}^{\text{median}}$ | $z_{\text{reion,H\,I}}^{\text{end}}$ | $\Delta z_{\text{reion,H\,I}}$ | $\tau_e$ | $\Delta T_{\text{H\,I}}$ [K] | $\eta$ | $M_{\text{min}}$ [M$_\odot$] |
|---|---|---|---|---|---|---|---|---|
| IR-A | Inhomogeneous (Model A) | 7.75 | 6.05 | 4.82 | 0.7120 (0.06260) | $2 \times 10^4$ | 11.0 | $1 \times 10^8$ |
| IR-A-noTthres | Inhomogeneous (Model A-no T threshold) | 7.75 | 6.05 | 4.82 | 0.7120 (0.06260) | $2 \times 10^4$ | 11.0 | $1 \times 10^8$ |
| IR-A-dz | Inhomogeneous (Model A-$\Delta z$) | 7.75 | 6.05 | 4.82 | 0.7120 (0.06260) | $2 \times 10^4$ | 11.0 | $1 \times 10^8$ |
| IR-A-li | Inhomogeneous (Model A-lin-interp) | 7.75 | 6.05 | 4.82 | 0.7120 (0.06260) | $2 \times 10^4$ | 11.0 | $1 \times 10^8$ |
| IR-A-gs | Inhomogeneous (Model A-gauss-smo) | 7.75 | 6.05 | 4.82 | 0.7120 (0.06260) | $2 \times 10^4$ | 11.0 | $1 \times 10^8$ |

Column 1: Simulation code.
Column 2: Method used to simulate reionization. See text for more details.
Column 3: H I reionization median redshift.
Column 4: H I reionization end redshift.
Column 5: Width of H I reionization. $\Delta z_{\text{reion,H\,I}} = z_{\text{reion,H\,I}}^{0.99} - z_{\text{reion,H\,I}}^{0.1}$.
Column 6: Thompson scattering optical depth $\tau_e$ density-weighted (volume-weighted).
Column 7: Total heating assumed for H I reionization.
Column 8: Parameter excursion set model 1. Efficiency.
Column 9: Parameter excursion set model 2. Minimum mass.
The simulations used for these tests have a box size of length, $L_{\text{box}} = 20$ Mpc/$h$ and $1024^3$ resolution elements.

this result is not relevant for the results presented in this work, it raises some doubts about the accuracy at very small scales $k > 0.1$ s km$^{-1}$. It is, however, still not clear if these differences are caused by the spatial resolution of the hydrodynamical simulation or by some intrinsic limitations of the reionization models used in this work. Recent works have shown that pushing observations of the 1D flux power spectrum beyond the classical limit of $k = 0.1$ s km$^{-1}$ could significantly improve the various constraints obtained from this observable on H I reionization and probably other physics (e.g., the nature of dark matter; Nasir et al. 2016). Therefore a detailed analysis of these high-resolution effects will be crucial for accurate models of the 1D flux power spectrum.

MNRAS **000**, 1–22 (2019)

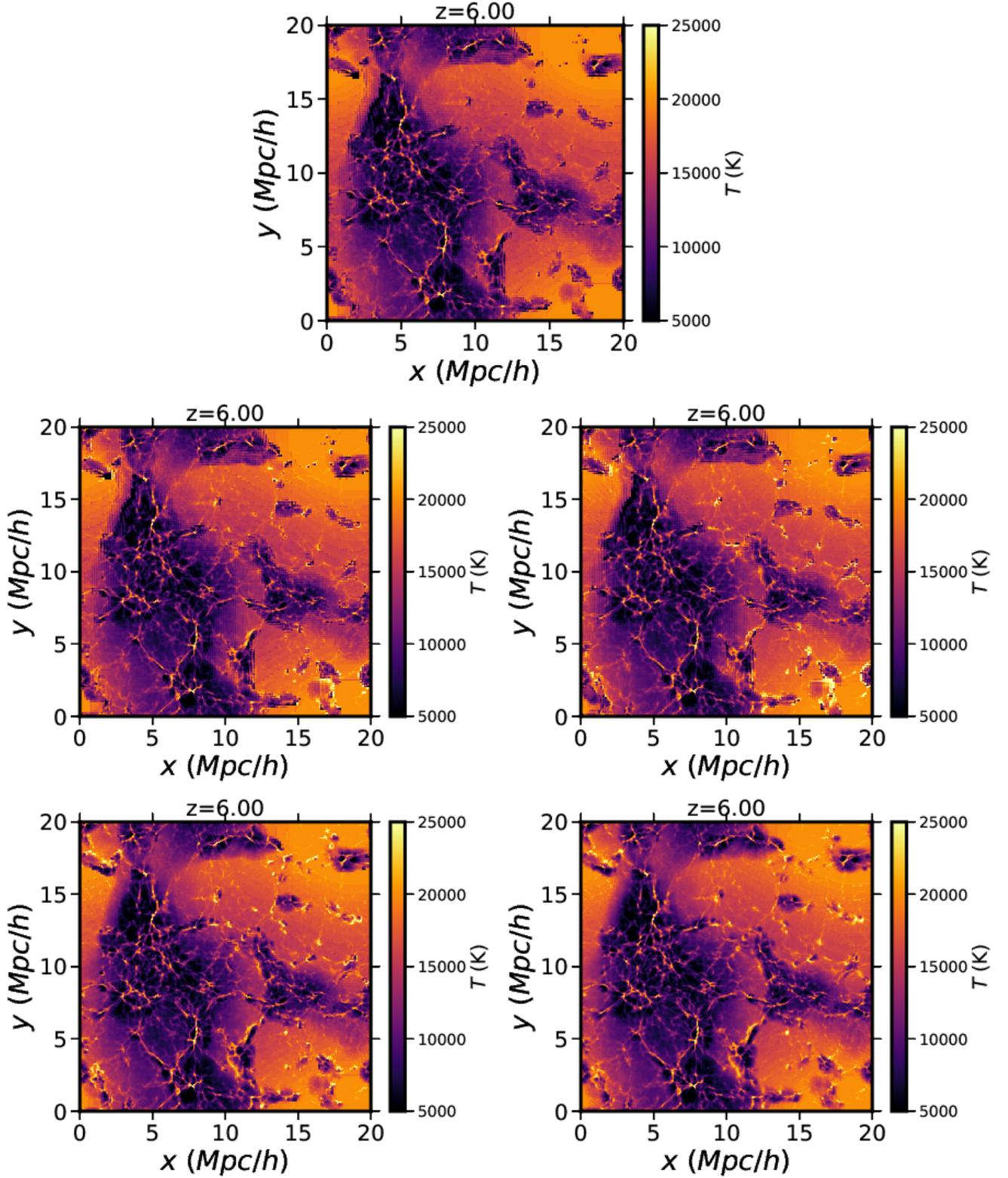

**Figure A1.** Slice of the temperature field at $z = 6$ for various simulation tests on the small-scale structure of the reionization field. Top row: method used in our fiducial runs. Second row left panel: reionization heating in the simulation with no temperature threshold. Second row right panel: input reionization field generated with a higher time resolution. Bottom row left panel: input reionization field smoothed using linear interpolation. Bottom row right panel: input reionization field smoothed using gaussian interpolation. See text for more details.

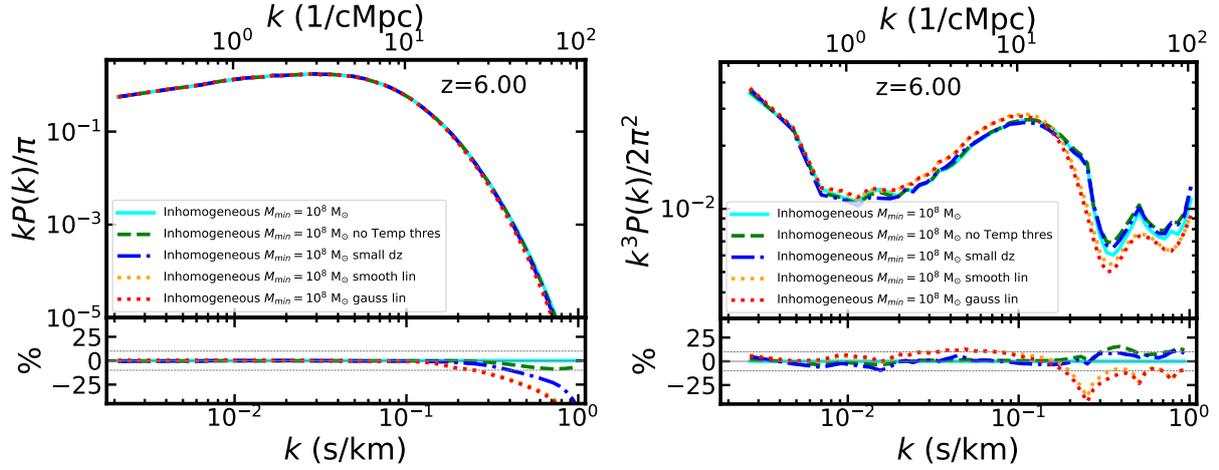

**Figure A2.** The 1D flux power spectrum at $z = 6$ (left panel) and the 3D temperature power spectrum at $z = 6$ (right panel) for a set of simulation tests on our new method to simulate inhomogeneous reionization models in hydrodynamical simulations. Both plots show the following models: our fiducial method (cyan solid line, IR-A), reionization heating in the simulation with no temperature threshold (dashed green line, IR-A-noTthres), input reionization field generated with a higher time resolution (dot-dashed blue line, IR-A-dz), input reionization field smoothed using linear interpolation (dotted orange line, IR-A-li), input reionization field smoothed using gaussian interpolation (red dotted line, IR-A-gs). We do not find any significant changes in the main results presented in this work for any of these variations. See text for more details.